\PassOptionsToPackage{dvipsnames}{xcolor}
\documentclass[%
 reprint,
 amsmath,amssymb,
 aps,superscriptaddress
]{revtex4-2}

\usepackage[printonlyused]{acronym}
\usepackage{float}
\usepackage{mathtools} 
\usepackage{extarrows} 
\usepackage{graphicx}
\usepackage{subcaption}
\usepackage{dcolumn}
\usepackage{bm}
\usepackage{makecell}
\usepackage{array}
\usepackage{rotating}
\usepackage[hidelinks]{hyperref}
\hypersetup{breaklinks=true}
\usepackage{microtype}
\usepackage{orcidlink}
\usepackage{tabularx}
\usepackage{booktabs}
\usepackage{braket}
\usepackage{svg}
\usepackage{siunitx}
\usepackage[justification=justified]{caption}
\usepackage[english]{babel}
\usepackage[most]{tcolorbox}
\newcolumntype{C}{>{\centering\arraybackslash}X}

\usepackage{endnotes}

\definecolor{d-blue}{HTML}{042c58}

\begin{document}

\preprint{APS/123-QED}

\title[Visualization enhances multi-Qubit Problem Solving]{Visualization Enhances Problem Solving in multi‑Qubit Systems}

\author{Jonas Bley}
\email{correspondence: jonas.bley@rptu.de}
\affiliation{Department of Physics and Research Center OPTIMAS, RPTU University Kaiserslautern-Landau, Erwin-Schroedinger-Str. 46, 67663 Kaiserslautern, Germany}
\affiliation{%
 Embedded Intelligence, German Research Center for Artificial Intelligence, 67663 Kaiserslautern, Germany
 }%

\author{Eva Rexigel}
\affiliation{Department of Physics and Research Center OPTIMAS, RPTU University Kaiserslautern-Landau, Erwin-Schroedinger-Str. 46, 67663 Kaiserslautern, Germany}

\author{Alda Arias}
\affiliation{Department of Physics and Research Center OPTIMAS, RPTU University Kaiserslautern-Landau, Erwin-Schroedinger-Str. 46, 67663 Kaiserslautern, Germany}%

\author{Lars Krupp}
\affiliation{%
 Embedded Intelligence, German Research Center for Artificial Intelligence, 67663 Kaiserslautern, Germany
 }%
\affiliation{Department of Computer Science and Research Initiative QC-AI, 67663 Kaiserslautern, Germany}

\author{Nikolas Longen}
\affiliation{Department of Computer Science and Research Initiative QC-AI, 67663 Kaiserslautern, Germany}

\author{Paul Lukowicz}
\affiliation{%
 Embedded Intelligence, German Research Center for Artificial Intelligence, 67663 Kaiserslautern, Germany
 }%
\affiliation{Department of Computer Science and Research Initiative QC-AI, 67663 Kaiserslautern, Germany}

\author{Stefan Küchemann}
\affiliation{Faculty of Physics, Chair of Physics Education, Ludwig-Maximilians-Universität München (LMU Munich), 80539 Munich, Germany}

\author{Jochen Kuhn}
\affiliation{Faculty of Physics, Chair of Physics Education, Ludwig-Maximilians-Universität München (LMU Munich), 80539 Munich, Germany}

\author{Maximilian Kiefer-Emmanouilidis}
\affiliation{%
 Embedded Intelligence, German Research Center for Artificial Intelligence, 67663 Kaiserslautern, Germany
 }%
\affiliation{Department of Computer Science and Research Initiative QC-AI, 67663 Kaiserslautern, Germany}

\author{Artur Widera}
\affiliation{Department of Physics and Research Center OPTIMAS, RPTU University Kaiserslautern-Landau, Erwin-Schroedinger-Str. 46, 67663 Kaiserslautern, Germany}

\begin{abstract}
Quantum Information Science (QIS) is a vast, diverse, and abstract field. In consequence, learners face many challenges. Science, Technology, Engineering, and Mathematics (STEM) education research has found that visualizations are valuable to aid learners in complex matters. The conditions under which visualizations pose benefits are largely unexplored in QIS education. In this eye-tracking study, we examine the conditions under which the visualization of multi-qubit systems with the Dimensional Circle Notation (DCN) in addition to the mathematical symbolic Dirac Notation (DN) is associated with a benefit for solving problems on the ubiquitously used Hadamard gate operation in terms of performance, Extraneous Cognitive Load (ECL) and Intrinsic Cognitive Load (ICL). We find that DCN increases performance and reduces cognitive load for participants with little experience in quantum physics. In addition, representational competence is able to predict reductions in ECL with DCN, but not performance or ICL. Analysis of the eye-tracking results indicates that task solvers with more transitions between DN and DCN benefit less from the visualization. We discuss the generalizability of the results and practical implications. 
\end{abstract}

\keywords{Visualization, multi-Qubit Systems, Multimedia Testing, Cognitive Load theory, Eye-Tracking, Integrated Theory of Text and Picture Comprehension, Representational Competence}

\maketitle

\section{Introduction}\label{sec:intro}

\subsection{Background and Motivation}

The two-level quantum system, also known as a qubit, is at the core of the field of \ac{qis}, where the concepts of superposition and entanglement find and raise further hopes for multidisciplinary applications ranging from measuring time~\cite{Nichol2022}, magnetic fields~\cite{Aslam2023} and gravitation~\cite{RademacherMillenLi,Stray2022}, simulating complex systems~\cite{Daley2022}, to secure communication~\cite{Couteau2023}, information processing and computing~\cite{Blais2020,Huang2020} and machine learning~\cite{Cerezo2022}. In many of these areas of application, systems of multiple qubits offer advantages over systems of single qubits. In \ac{qis}, entangled multi-qubit superposition states are used for technological advantages. These systems cannot be represented by describing every qubit individually, but have to be represented as a whole, if one does not wish to lose information. 

The Hadamard gate given by the rules H$\ket{0}=1/\sqrt{2} (\ket{0}+\ket{1})=\ket{+}$ and H$\ket{1}=1/\sqrt{2} (\ket{0}-\ket{1})=\ket{-}$, using the \ac{dn} is a typical operation for creating or destroying superpositions and is therefore of central importance for the field of \ac{qis}. Using the Hadamard gate, quantum states can be moved from the computational basis consisting of the basis states $\ket{0}$ and $\ket{1}$ to the Hadamard basis consisting of the basis states $\ket{+}$ and $\ket{-}$. This opens the computational opportunity of operating on the phases of qubits, i.e., the relative phase difference between the $\ket{0}$ and $\ket{1}$ states. This is a central distinction of quantum computing from classical computing that finds application in all pillars of quantum technologies. An example of a common concept that uses the Hadamard gate is phase kickback, where the role of control and target qubit is reversed by kicking the phase of the target qubit back to the control qubit. This kick-backed phase can then be measured in the Hadamard basis~\cite{Wang_2023}. In this way, the Hadamard gate can be used to extract phase information from a system of qubits and, therefore, enable oracle-based quantum algorithms like Quantum Fourier Transform, Quantum Phase Estimation, the Deutsch-Josza algorithm, Grover's algorithm, and quantum error correction protocols~\cite{Fundamentals}. 

In recent years, the number of educational \ac{qis} offerings has seen a large increase, ranging from standalone introductory courses to full bachelor's and master's programs~\cite{Aiello_2021,10.1117/1.OE.61.8.081806}, spurred by initiatives such as the US National Quantum Initiative~\cite{NQIA2018} or the European Quantum Flagship~\cite{EUQuantumFlagship2018}. This rapid growth has produced a patchwork of curricula and teaching approaches, underscoring an urgent need to investigate and develop evidence-based teaching methods within \ac{qis}~\cite{PhysRevPhysEducRes.20.010131,PhysRevPhysEducRes.20.020601}. This need is further underlined by the typical challenges that learners face in the field~\cite{Nita03042023,PhysRevPhysEducRes.20.020108,Seifollahi_2025}.

In previous research, many difficulties have been identified in \ac{qis} teaching and learning. In interviews, student difficulties with \ac{dn}, phase kickback and the Toffoli and rotation gates became apparent~\cite{10313606}. Similarly, in~\cite{PhysRevPhysEducRes.20.020108}, students were shown to have difficulties with measurement outcomes, transitioning between the bra ($\bra{\psi}$) and the ket ($\ket{\psi}$) states, using outer products ($\ket{\psi}\bra{\psi}$) and \ac{dn} in general. It is of great importance to answer the question of how to alleviate such issues that students have when learning the basics of quantum computing. A known way of supporting students in \ac{stem} education, especially in dealing with complex content, is the effective use of \ac{mer}~\cite{AINSWORTH1999131}. In a pilot study, we identified the Hadamard gate in multi-qubit systems to be an interesting avenue for further exploration when considering the benefits of the \ac{cn} and \ac{dcn} visualization in task solving~\cite{bley2025visualizingquantumstatespilot}.

Meta analyses have shown that there are also benefits to using \ac{mer} in task solving as opposed to learning, while more primary research is also necessary in this domain~\cite{Hu2021,Schewior2024}. This research should serve the purpose of determining the influence of possible moderators, such as students' previous knowledge, item complexity, \acf{rc}, and eye movements. Here, we discuss the relations between these theoretical concepts using novel methods and a combination of known methods. We therefore take an important step in the domain of educational research in \ac{stem} more broadly.

\subsection{Visual and cognitive processes in task solving with MERs}

Learning in general is defined as changes of long-term memory \cite{Paas_Sweller_2014}, also referred to as schema acquisition~\cite{mayer2005}. The \ac{ctml} is a framework explaining how this takes place. It assumes that learners make use of two channels, one visual and one auditory, in order to receive external information (the dual channels assumption~\cite{mayer2005}). Although written text and pictures are both received in the same cognitive channel, they can still be processed in two different parts of the working memory. For this reason, learning with both text and pictures can be more effective than learning with text alone, which is known as the multimedia principle~\cite{fletcher2005multimedia}. 

Task solving refers to completing specific problems with a defined goal. As we specifically evaluate task-solving performance, it is here placed within the theory of Multimedia Testing (MMT), as opposed to the more broad theory of Multimedia Assessment (MMA) \cite{Kirschner2017}. As learning and task solving share underlying cognitive processes, representations can be evaluated in the context of task solving to produce statements about learning~\cite{Schewior2024}. However, while task solving and learning share the need to construct mental schemata, the step unique to task solving is the decision-making step~\cite{https://doi.org/10.1002/acp.3060,LINDNER201791}. The results of Lindner's studies show that students with a higher level of understanding will dismiss incorrect answer options more readily, leading to the conclusion that, if pictures enhance the level of understanding, gaze time spent on the correct options should be increased in relation to the time spent on the text. However, such effects have been shown to not be unconditional. For example, purely decorative pictures have been shown to have no impact on performance~\cite{LINDNER2020101345,Schewior2024} and pictures that are both interesting and irrelevant, also known as seductive details, have even been shown to hinder learning~\cite{Sundararajan2020}. In task solving, there is evidence of beneficial effects of representational pictures in certain contexts, under conditions that could include prior knowledge, age, and affective motivational factors~\cite{Schewior2024}. In this paper, we refer to representational pictures as visualizations, demarcating them from mathematical symbolism and decorative pictures. There is a large amount of evidence that visualization can provide benefits for learners in \ac{stem} under the right circumstances, that is, if the representation supports relevant cognitive processes (outlined in this section) and if the learners possess \acf{rc}~\cite{gilbert_visualization_2005,Munfaridah2021}. The latter is discussed in Section \ref{sec:rep_comp_sr}.

The \ac{ctml} is based on two main assumptions: First, human working memory is limited, which means that too much information to be processed at once can lead to cognitive overload. As redundant information can add to working memory usage, it should be added with care. Secondly, it is assumed that new information is not only processed, but also actively integrated into a mental representation that is connected to previous knowledge and mental schemata, a process that facilitates learning~\cite{mayer2005}. 

The process model of multimedia learning goes into more detail~\cite{stark2018emotional}. It divides cognitive processes into five processes that eye-tracking data can be matched to as follows~\cite{stark2018emotional,https://doi.org/10.1111/jcal.13051}:

\begin{enumerate}
    \item The selection of the relevant verbal and pictorial information: \textit{number of fixations}
    \item The organization of the information within working memory: \textit{fixation duration, number of transitions and visit duration after transitions}
    \item Elaboration, i.e., integration of the information with prior knowledge: \textit{mean fixation duration}
    \item Metacognitive processes like planning, monitoring and regulating: \textit{number of transitions with short subsequent visit duration}
    \item Extraneous processes that are irrelevant for achieving the objective: \textit{percentage of irrelevant fixation time}
\end{enumerate}

Therefore, the number of fixations or fixation duration is used to describe visual attention and engagement in processing the information that takes place in all processes. Transitions between the corresponding areas of interest have been associated with greater engagement in the organization of the given information, especially in horizontal coherence formation, i.e., the process of matching the pictorial with the verbal information, and metacognitive processes~\cite{stark2018emotional,https://doi.org/10.1111/jcal.13051}. In the eye-tracking analysis within this work, we focus on process 2 (organization) to analyze intrinsic cognitive processes used for task solving that potentially influence performance.

\subsection{Functions, Tasks and Design of multiple external representations}\label{sec:functions}

The combination of symbolic and graphical external representations can yield various benefits in \ac{stem} education~\cite{Ainsworth2021, Hu2021}. However, as already discussed, there are certain conditions for their effectiveness. Ainsworth asked the question \textit{why} there can be benefits to \ac{mer} and formulated the functions of \acf{mer}~\cite{AINSWORTH1999131}, culminating in the \ac{deft}-framework~\cite{ainsworth2006}. This framework includes the functions that the representations fulfill, the design of the presentation of the representations, as well as the associated cognitive tasks. The possible functions of \ac{mer} include complementing each other by supporting different cognitive processes or information, constraining interpretation by familiarity or inherent processes, and constructing deeper understanding by abstraction, relation or extension. While \ac{mer} can be redundant in the information they provide, they can still support complementary processes and, therefore, can lead to benefits for learners and task solvers. Beyond the functions discussed by the \ac{deft}-framework, certain visualizations have also been shown to be capable of increasing student self-confidence~\cite{novriani2019development} and motivation~\cite{7817884} and letting them experience less negative and more positive emotions associated with learning~\cite{LACAVE2020103817}.

According to the \ac{deft} framework, the cognitive tasks that learners undertake are due to the characteristics of the external representations, as well as the learner characteristics. Representational characteristics include, but are not limited to, the sensory channels (visual and/or auditory channels) that the representations use, the modality (e.g., combining symbols or text with graphical representations), the level of abstraction, the type (e.g., equations, tables, line graphs, different types of texts, etc.), integrated presentation of \ac{mer}, whether the representations are dynamic or static, and the dimensionality. Learner characteristics include familiarity with the representations or the domain, the age, and other individual differences such as \ac{sra}. Cognitive tasks are imposed on the learner by the design of the \ac{mer} and the characteristics of the representations. Cognitive tasks, as described by the \ac{clt}, include the usage of working memory for tasks that are useful to the goal to be obtained, also called \ac{icl}, the usage of working memory for tasks that are not essential for the goal, also called \ac{ecl}, and the use of working memory to organize selected information into mental schemata and integrate it into previous knowledge, also known as \ac{gcl}. \ac{gcl} is therefore also referred to as the cognitive load required for learning processes~\cite{SWELLER1988257,Mayer_2021}. However, it can be argued that \ac{gcl} and \ac{icl} are redundant concepts that could be merged~\cite{kalyuga2011cognitive}, but it is an ongoing topic of debate and results tend to support a three-split model of cognitive load~\cite{Krieglstein2022}.

The level of expertise can have a significant impact on the efficacy of \ac{mer}. The redundancy effect suggests that informationally redundant material hinders learning due to claiming working memory load, increasing \ac{ecl}~\cite{KalyugaSweller2014}. However, when novices lack the required ability to extract information from one of the representations, they can benefit from the use of \ac{mer} in cases where experts might not benefit. For experts, this might only result in increased \ac{ecl} and therefore lower performance. This effect is called the expertise-reversal effect~\cite{Sweller2011,ReyBuchwald2011}. 

There are certain design principles for the presentation of \ac{mer} that can be followed to reduce \ac{ecl}, leaving more room for allocation of working memory for useful tasks~\cite{Mayer01012003,mayer201412}. These design principles include coherence (avoiding seductive details), signaling (highlighting key content), avoiding redundancy, and spatial and temporal contingency (avoiding separation of \ac{mer})~\cite{MAYER2021229}. While it is useful to follow such design principles, it is also essential to consider how those parts of the working memory that serve essential purposes can be aided. This is given if the presentation of \ac{mer} reduces \ac{icl} of the same task. 

\subsection{\acl{rc} and Spatial Reasoning}\label{sec:rep_comp_sr}

\acf{rc} is an important condition for the effectiveness of \ac{mer}~\cite{rau2017conditions}. It refers to knowledge of how visualizations represent information and the skills to apply this knowledge~\cite{rau2017conditions}. In the case of qubit systems, this information can be the underlying nature of the depicted quantum state, like the entanglement properties, or a given process, such as transitioning between two quantum states with the Hadamard gate. It is important to note that \ac{rc} is dependent not only on the learner and the visualization, but also on the given context in which the visualization is used. The same visualization can serve different functions according to the \ac{deft} framework, depending on the context to which it is applied. \ac{rc} is, by definition, necessary for learners to benefit from visualizations~\cite{rau2017conditions}. Rau divides \ac{rc} into three categories: conceptual competencies, perceptual competencies, and meta-representational competencies, which are described below.

Conceptual competencies are separated into visual understanding and connectional understanding. Visual understanding consists of the ability to identify relevant visual features and connect the visualization to concepts, symbolism, or text. It also includes general principles, conventions, and the ability to communicate using visualization. In this work, we include the ability to translate between symbolic and visual notations (translation competence) and to understand the action of the Hadamard gate visually (procedural competence) in the concept of visual understanding. Connectional understanding is the ability to connect multiple visualizations with one another, identify relevant similarities, and communicate and understand conventions around these processes. Perceptual competencies include visual and connectional fluency. Visual fluency is regarded as the efficiency in connecting visualizations to concepts and chunking (grouping together of visual features to form conceptual segments, which allows for more efficient cognitive processes) and communication about these processes. Connectional fluency involves the efficiency of connecting multiple visualizations and multiple chunks within different visualizations to each other, and the flexibility of switching between representations. Meta-representational competencies consist of the ability to choose appropriate visualizations based on task demands, contexts, own ability level, and personal goals. In scenarios where only a single visualization is used, connectional fluency and understanding are not involved in the learning process, and meta-representational competencies are less important~\cite{rau2017conditions}. 

To understand the cognitive processes that explain how visualizations are used in more detail, it is necessary to consider the \ac{itpc}~\cite{schnotz2005}. The \ac{itpc} goes beyond the \ac{ctml} to describe the sequential process of the construction of internal representations within the auditory and the visual working memory, after visual and/or auditory information has entered the sensory register. The model divides visual information into graphemic/symbolic (i.e., text, symbolism, formulas, etc.) and visuo-spatial (pictures, photographs, drawings, etc.) inputs and acoustic information into non-verbal and phonological inputs. During this process of organization to form internal patterns, visual symbolic information is converted to phonological information (grapheme-phoneme conversion), such that both working memory channels are utilized. After input and feature analysis of the presented information, the \ac{itpc} divides the processing in the working memory into depictive and descriptive processing, the former resulting in an internal, depictive mental model and the latter resulting in a propositional, more functional internal representation. The mental model and the propositional representations are more conceptual than the previously organized visuo-spatial and symbolic patterns. The two internal representations are evaluated and compared, where the propositional representation can be inspected using the mental model, and the mental model can be constructed or updated using the propositional representation. For example, the action of a Hadamard gate could be constructed in a propositional form, $H\ket{0}\to \frac{1}{\sqrt{2}}(\ket{0}+\ket{1})$ and $H\ket{1}\to \frac{1}{\sqrt{2}}(\ket{0}-\ket{1})$, or it could be seen as splitting or combination of circles, with phase flips corresponding to movement to the left (see Section \ref{sec:dn_and_cn}). Comparing the two models for coherence formation can be a deliberate or subconscious process~\cite{10.3389/fpsyg.2017.00827}. Here, it is used evaluate and compare the two representations of the action of a Hadamard gate in order to solve the given task.

More functional thinking is an indication of expert thinking~\cite{10.3389/feduc.2023.1192708}. When embedding Rau's concept of visual understanding into the \ac{itpc} and to our case, we see the procedural competence described in the previous paragraph as aiding in depictive processing, while translational competence aids in model inspection. In the example of the Hadamard gate, the action of the quantum gate can be understood in the descriptive \acf{dn} and/or in the depictive \acl{dcn}, the latter requiring visual understanding of these processes that we call procedural competence. If only \ac{dcn} is understood but not the functioning of the Hadamard gate, learners can translate to \ac{dn} and make use of this aspect of visual understanding, which we call translational competence.

\acf{sra} can be especially helpful in the early stages of learning a concept and strongly predicts success in \ac{stem} fields~\cite{UTTAL2012147}. In the \ac{itpc}, spatial reasoning is seen as part of the depictive mental model construction, but can also aid in the development of a functional propositional representation~\cite{schnotz2005}. Using representations that rely on \ac{sra} therefore aids in the construction of functional propositional representations from depictive mental models and increases success in \ac{stem} learning in general~\cite{Taylor2023}. One can distinguish different types of spatial thinking abilities between reference frame (intrinsic or extrinsic), small scale and large scale, activity level (static or dynamic), and dimensionality (2D or 3D). For example, the ability to mentally rotate objects uses an intrinsic reference frame (with no relation to another object) and a dynamic activity level. 

\begin{figure*}[ht]
    \centering
    \includegraphics[width=\linewidth]{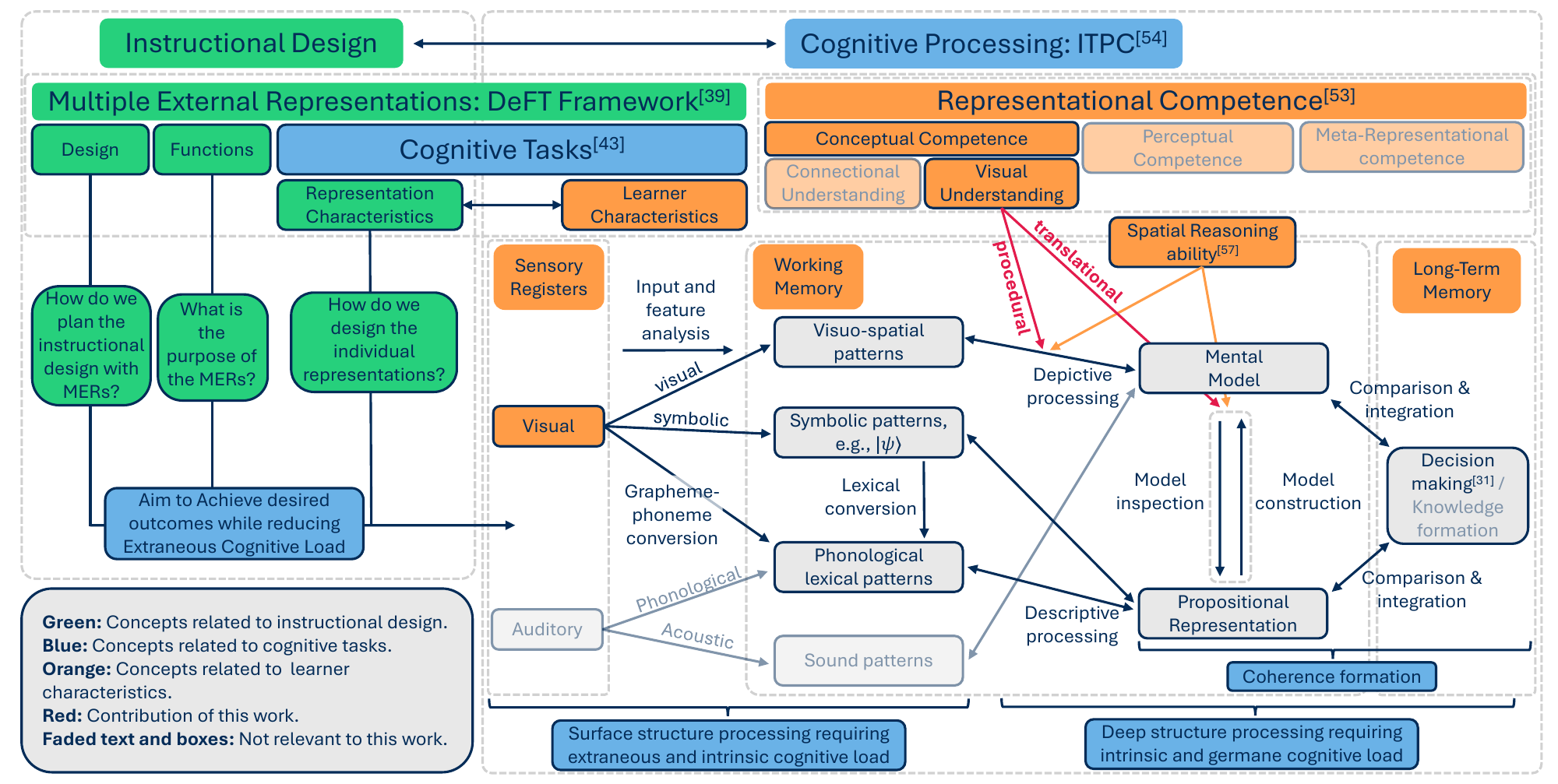}
    \caption{Interrelations between the \acf{deft}-framework and the \acf{itpc}. The applications of the Designs and Functions of the \ac{deft}-framework lie on the spectrum of instructional design of \acf{mer}. In this study, we start with visual information that is symbolic (\ac{dn}) or visual (\ac{dcn}) and focus on how this information is processed. Within the Tasks of the framework, learner characteristics such as \acf{rc} and \acf{sra} are considered. As part of \ac{rc}, we only consider visual understanding as we only use only a single visual representation~\cite{rau2017conditions}. We make theoretical hypotheses to embed the theory around \ac{rc} to the \ac{itpc}: Visual understanding and \ac{sra} can aid in the depictive mental model construction from visual patterns and in evaluation/inspection of the mental model from the descriptive propositional representation. In the former, what we call procedural competence is involved, while in the latter, translational competence plays a main role. In the end of the task solving process, the mental model and/or propositional representation are compared to the candidate solutions to come to a decision.}
    \label{fig:schema_theory}
\end{figure*}

Figure \ref{fig:schema_theory} shows an overview of the theories around \ac{mer} involved in this work: \ac{deft} and the \ac{itpc}, and how they relate to the \acl{clt}, \ac{rc}, and \ac{sra}. In the next section, we will discuss how different representations used in \ac{qis} can utilize the underlying concepts of the \ac{itpc} to enhance learning and task solving.

\subsection{Dirac and Circle Notation for QIST education}\label{sec:dn_and_cn}

A quantum state in the computational basis can be written as 

\begin{equation}
    \ket{\psi}=\sum_{i\in\{0,1\}^n} \alpha_i\ket{i}.
\end{equation}

\begin{figure}[ht]
    \centering
    \includegraphics[width=0.5\linewidth]{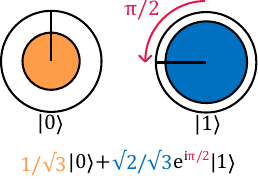}
    \caption{A single-qubit superposition state in \acl{cn}. The radius of the inner circles are the absolute values of the complex numbers (here, $1/\sqrt{3}\approx 0.58$ and $\sqrt{2}/\sqrt{3}\approx0.82$), and the phase is represented by the angle of the line starting from the vertical position counterclockwise (here, 0 on the left and $\pi/2$ on the right).}
    \label{fig:circ_not}
\end{figure}

The \acf{dn} is, as the main language of \ac{qis}, a relevant part of \ac{qis} education~\cite{rexigel2024investigating}. It also has some didactical advantages regarding sense-making of fundamental physical principles and the connection to the underlying mathematical ideas~\cite{PhysRevPhysEducRes.20.020134}. The \ac{dn} comes with two major hurdles to overcome: First, complex numbers can be intimidating to learners new to the field. The \ac{cn} has been proposed to alleviate this issue~\cite{johnston_harrigan_gimeno-segovia_2019}. In \ac{cn}, each basis state is associated with a circle, and the amplitudes of the complex coefficients of the basis states are depicted as an inner circle describing their magnitude and a gauge line describing their complex phase. This is shown in Figure \ref{fig:circ_not}.

\begin{figure*}[ht]
    \centering
    \includegraphics[width=0.8\linewidth]{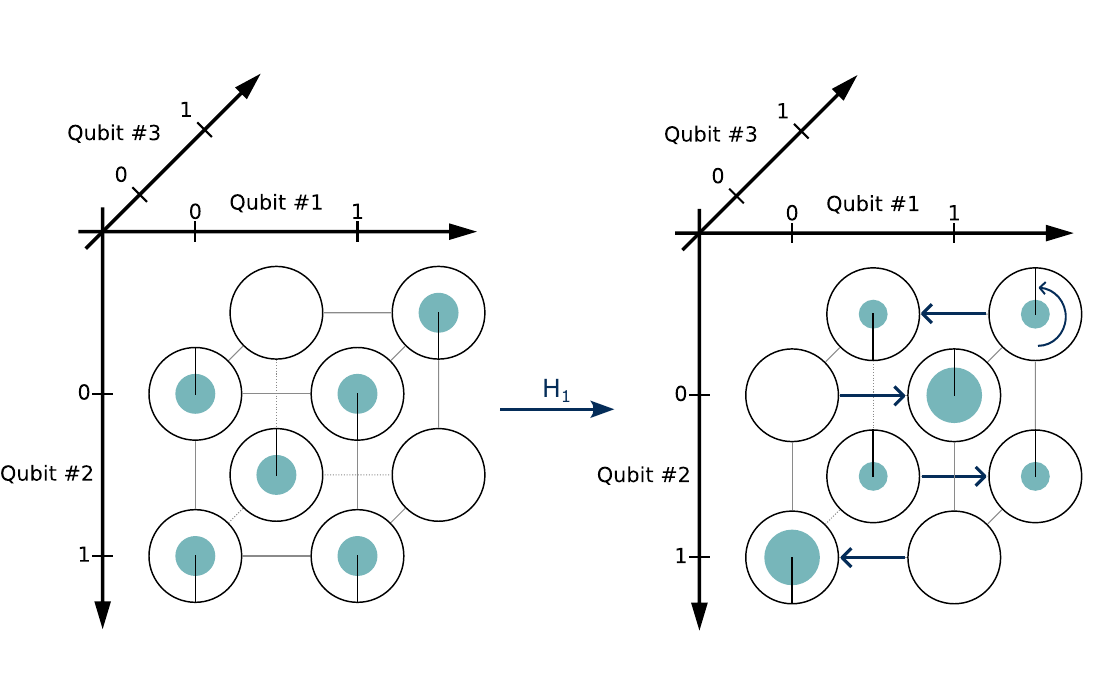}
\caption[Action of the Hadamard gate in DCN.]{Action of the Hadamard gate in \ac{dcn}. The $H_1$ gate is applied to the state $\frac{1}{\sqrt{3}}\ket{001} - \frac{1}{\sqrt{3}}\ket{011} - \frac{1}{2\sqrt{3}}\ket{100} + \frac{1}{2\sqrt{3}}\ket{101} + \frac{1}{2\sqrt{3}}\ket{110} + \frac{1}{2\sqrt{3}}\ket{111}$ to obtain $\frac{1}{\sqrt{6}}\ket{000} - \frac{1}{\sqrt{6}}\ket{001} - \frac{1}{\sqrt{6}}\ket{010} + \frac{1}{\sqrt{6}}\ket{011} - \frac{1}{\sqrt{6}}\ket{101} + \frac{1}{\sqrt{6}}\ket{110}$ (with basis states written as $\ket{\text{qubit} \#3 \text{ qubit} \#2 \text{ qubit} \#1 }$), following the rules $\ket{0}\xleftrightarrow{H}1/\sqrt{2} (\ket{0}+\ket{1})$ and $\ket{1}\xleftrightarrow{H}1/\sqrt{2} (\ket{0}-\ket{1})$ along the axis of qubit \#1. The four resulting combination of these rules are depicted in the figure.}
    \label{fig:hadamard}
\end{figure*}

Second, when representing systems of multiple qubits in \ac{dn}, the role of one qubit to the whole system can become unclear as one qubit is part of every single basis state. This means that properties like (partial) entanglement can remain hidden, but also that the action of quantum operations on specific qubits can become unintuitive as they are applied on different combinations of basis states, depending on the qubit(s) involved and the ordering of the states. Combining the visual depiction of complex phases and assignment of qubits to axes in space was introduced in~\cite{just_2021} and was shown to visualize entanglement properties using the extension of \acf{cn} to \acf{dcn}~\cite{bley2024visualizing}. This approach can make use of the cognitive capacity for spatial reasoning to convey the fundamental concepts of quantum computing. As described by the \ac{itpc} (see Section \ref{sec:rep_comp_sr}), the spatial depiction of qubits could help learners build more functional internal representations. Figure \ref{fig:hadamard} shows the action of the Hadamard gate in \ac{dcn}. As can be seen there, the Hadamard gate acts along the axis of the qubit it is imposed on. In this way, the visualization offers a geometric strategy to describe the process. 

In the \ac{itpc}, the \ac{dn} is a descriptive representation of quantum states, while \ac{dcn} is depictive. Within the framework of \ac{rc}, we can differentiate between understanding the visualization of static quantum states (and measure this, e.g., by assessing the ability of translating between \ac{dn} and \ac{dcn}) and procedural competence with visualization (visually understanding the action of, for example, the Hadamard gate). Especially when used to represent quantum operations like the Hadamard gate, the added spatial information in \ac{dcn} can aid with the mental model construction of the operation that can then be used to apply the Hadamard gate correctly. The splitting/merging of amplitudes and flipping of phases is a visuo-spatial representation that can also be seen in \ac{cn}, while \ac{dn} supports only symbolic patterns. Whether and under which circumstances the visuo-spatial processing in addition to or instead of the symbolic processing can lead to increases in performance is, especially within the \ac{qis} context, an open research question.

\subsection{Research Questions and Hypotheses}\label{subsec:rq}

While a possible benefit of visualizations such as \ac{dcn} for learning is predicted by previous research, it is uncertain to what extent this benefit manifests also in task solving. Although there is evidence that visualizations also provide benefits in this realm, they are likely to depend on visualization, context, and task solver and the exact moderating factors have not yet been explored sufficiently~\cite{Schewior2024}. Possible factors include the previous experience of the student, which could result in less reliance on a visualization, \acf{sra} as described by the \ac{itpc}~\cite{schnotz2005}, and more particularly \acf{mra}~\cite{Taylor2023}, and \acf{rc}~\cite{rau2017conditions}, whereas possible differences may be due to support of the decision-making step~\cite{LINDNER201791}.

There is a need for studies considering also other moderators like learner and representational characteristics and eye-tracking data to further investigate the underlying processes and conditions for benefits of presenting a visualization in terms of accuracy of performance and perceived cognitive load. As \ac{gcl} is mainly associated with learning processes and we are in the realm of task solving, we choose \ac{icl} and \ac{ecl} as main outcome variables for cognitive load. As is apparent, the \ac{dcn} visualization adds a complementary strategy (as seen from the perspective of the \ac{deft} framework) to understanding the process of the application of a Hadamard gate in two- and three-qubit systems. The question of whether there is a benefit to this strategy and under which conditions remains to be answered. We formulate the following research questions.

\begin{description}
    \item[\textup{\textbf{RQ1}}] Do participants benefit from the visualization when solving questions on the Hadamard gate in two- and three-qubit systems in terms of\ldots\noindent
    \begin{description}
     \item[\textup{\textbf{a}}] \ldots accuracy?
     \item[\textup{\textbf{b}}] \ldots perceived cognitive load (\ac{icl} and \ac{ecl})?
    \end{description}
\end{description}

To evaluate our \ac{rc} test instrument, we ask the following research questions.

\begin{description}
    \item[\textup{\textbf{RQ2}}] Is our \ac{rc} test instrument able to predict\ldots
        \begin{description}
     \item[\textup{\textbf{a}}] \ldots accuracy increases?
     \item[\textup{\textbf{b}}] \ldots reductions in perceived cognitive load?
    \end{description}
\end{description}

We ask the following research questions to explore whether the integration of visualization with mathematics has an impact on task solving. 

\begin{description}
    \item[\textup{\textbf{RQ3}}] What influence do integration processes of the visualization with the mathematics have on\ldots
    \begin{description}
     \item[\textup{\textbf{a}}] \ldots accuracy?
     \item[\textup{\textbf{b}}] \ldots perceived cognitive load?
    \end{description}
\end{description}

By answering these research questions, we will be able to make statements about how to support task solving in the context of the Hadamard gate in multi-qubit systems. Subsequently, we discuss the generalizability of the results to other associated contexts in the realm of understanding multi-qubit system processes and conclude with recommendations for educators in the field of \ac{qis} as well as perspectives for future research.

\section{Methods}\label{sec:methods}

As we found the study design to be suitable, we adopted methodological paths similar to the pilot study we conducted~\cite{bley2025visualizingquantumstatespilot} with, however, multiple adaptations. We stay with the A-B crossover design and cognitive load measurements, but restrict the context of the questions to the Hadamard gate, employ eye-tracking techniques, introduce a new way of measuring \acf{rc} and also measure \acf{mra}.
 
\subsection{Participants}\label{subsec:participants}

The total number of participants in the study was 42, including 31 people who were relatively new to the field of quantum physics, with a maximum of two years' experience. Due to the possible influence of expertise-reversal effects, we excluded all participants with more than three years of experience in quantum physics from the analyses. Of the 31 remaining participants, 22 were male and nine female. The participants' ages ranged between 19-33 years. The main field of study was physics for ten participants, computer science or engineering for ten, other fields of natural sciences for five, and \ac{qis} for four. Furthermore, four of the participants were bachelor students, 14 students were in their master's degree, nine were doing their Ph.D., and one has a Ph.D.

All procedures performed in the study were in accordance with the ethical standards of the national research committee and with the Declaration of Helsinki of 1964 and its subsequent amendments. The study involved data collection, including eye movements, in an on-site survey that took about 60 minutes. Participation in this study was voluntary, anonymous, and under informed consent. The data collected included participants’ age, gender, field of study, quantum physics experience, and highest educational achievement. Personal data of participants is kept confidential and used solely for research purposes. The anonymized data supporting the findings of the study is openly available \cite{bley_2025_15382444}. 

\subsection{Study Design}\label{subsec:design}

For the present study, a within-subject design was chosen. This is, first and foremost, enabled by the possibility of designing multiple similar questions to test task solving in the same context, therefore being able to show some of these questions with and some without visualization, to make comparative assessments. 

\subsubsection{Study structure}

\begin{figure*}[ht]
    \centering
    \includegraphics[width=\linewidth]{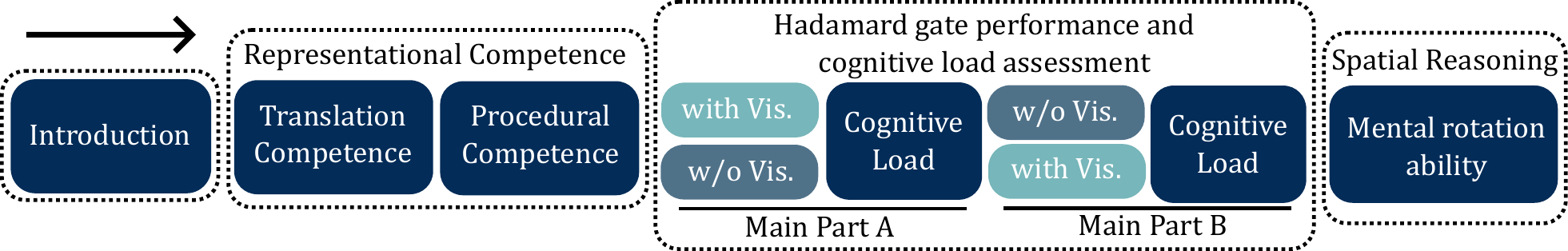}
\caption[Overall structure of the study.]{Overall structure of the study. Participants are divided into two groups: one is assigned questions with visualization first, the other without.}
    \label{fig:structure}
\end{figure*}

We first showed an introductory video explaining \ac{dcn} and the Hadamard gate to the participants. We then measured the translational understanding between \ac{dn} and \ac{dcn} with a set of eight questions and then the visual understanding of the Hadamard gate with a set of six questions. During the following main phase of the study, eye movements were tracked. During part A of the main phase, half of the participants were shown a block of six questions without visualization, and the other half of the participants were shown a block of questions with visualization. We then measured the perceived cognitive load during this part of the study, before giving each participant another block of six questions without or with visualization, depending on which type of question they did not receive during part A. This random order ensures that the learning effects during main part A of the study do not interfere with the results of main part B. Lastly, \acl{sra} was measured without eye-tracking. The complete structure of the study is shown in Figure \ref{fig:structure}.

\subsubsection{Introductory video}

The introductory video was used to decrease the influence of previous knowledge on the test results. Briefly, classical computing and Turing machines were mentioned, to motivate quantum computing and qubit states. Then, \ac{dcn} was introduced and the most important single-qubit operations were explained (X, Z, and H gates). To show operations in systems of multiple qubits, examples of the X and Z gates were shown in two-qubit systems, and the CNOT gate was shown in a three-qubit system. The video is shown in the supplementary material.

\subsubsection{Representational competence}

\begin{figure*}[ht]
\begin{tcolorbox}[colback=white, colframe=black]
Consider the following state.

Note: basis states are written as $|$qubit\#2 qubit \#1$\rangle$.

\includegraphics[width=0.3\linewidth]{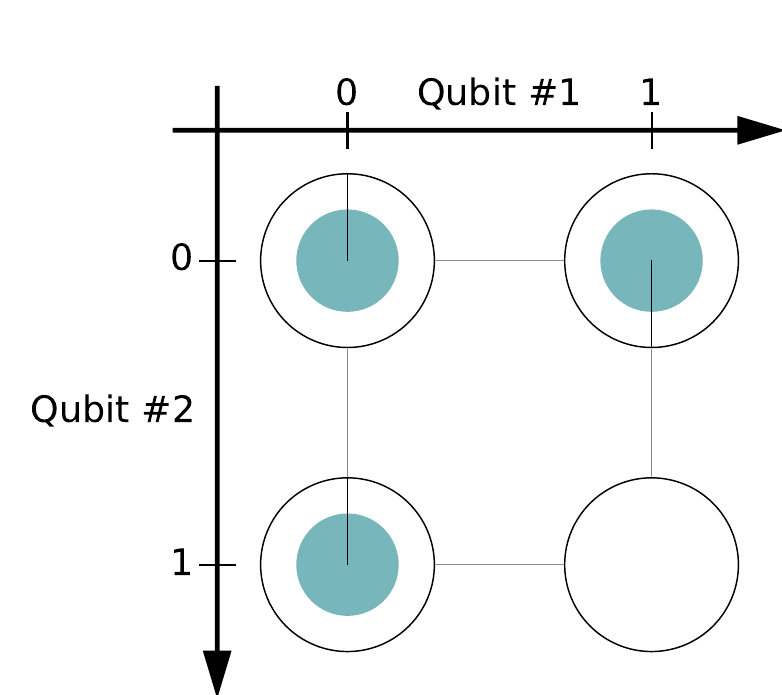}

Select the corresponding state in Dirac notation.
\begin{tcolorbox}[width=0.5\textwidth,boxsep=0pt,colback=white, colframe=d-blue, boxrule=0.5pt]
\begin{equation*}
    \frac{1}{\sqrt{3}}\ket{00}+\frac{1}{\sqrt{3}}\ket{01}-\frac{1}{\sqrt{3}}\ket{10}
\end{equation*}
\end{tcolorbox}
\begin{tcolorbox}[width=0.5\textwidth,boxsep=0pt,colback=white, colframe=d-blue, boxrule=0.5pt]
\begin{equation*}
    \frac{1}{\sqrt{3}}\ket{00}-\frac{1}{\sqrt{3}}\ket{10}+\frac{1}{\sqrt{3}}\ket{11}
\end{equation*}
\end{tcolorbox}
\begin{tcolorbox}[width=0.5\textwidth,boxsep=0pt,colback=white, colframe=d-blue, boxrule=0.5pt]
\begin{equation*}
    \frac{1}{\sqrt{3}}\ket{00}+\frac{1}{\sqrt{3}}\ket{10}-\frac{1}{\sqrt{3}}\ket{11}
\end{equation*}
\end{tcolorbox}
\begin{tcolorbox}[width=0.5\textwidth,boxsep=0pt, colback=white, colframe=d-blue, boxrule=0.5pt]
\begin{equation*}
    \frac{1}{\sqrt{3}}\ket{00}-\frac{1}{\sqrt{3}}\ket{01}+\frac{1}{\sqrt{3}}\ket{10}
\end{equation*}
\end{tcolorbox}
\end{tcolorbox}
\caption[Example of a translational understanding question.]{Example of a translational understanding question. The correct answer is at the bottom. The distractors are constructed by varying the position of the minus sign and the basis states from $\ket{01}$ and $\ket{10}$ to $\ket{10}$ and $\ket{11}$.}
\label{fig:tr_comp}
\end{figure*}

To stay as close as possible to the characterization of \acf{rc} provided by Rau~\cite{rau2017conditions}, but within feasibility, we decided to propose a test instrument to measure visual understanding as a measure of \ac{rc}, as the other types of representational competencies (connectional understanding and fluency, and meta-representational competencies) are only relevant when using more than one visualization. As mentioned at the end of Section \ref{sec:rep_comp_sr}, we differentiate between two types of relevant visual understanding constructs: translation competence and procedural competence for which we constructed two different types of questions. In the first type, participants were asked all eight combinations of translating between amplitudes or phases in systems of two or three qubits, from \ac{dcn} to \ac{dn} or back. An example is shown in Figure \ref{fig:tr_comp}. 

The second type of \ac{rc} question asked about understanding the action of the Hadamard gate in \ac{dcn} (procedural competence). Here, the Hadamard gate was explained in text format and the participants' ability to transfer this information from the text to the action of the Hadamard gate in the visualization was tested. We created one question on solely creating superpositions, one on destroying them, and one on a mixture of both, for systems of two and three qubits. The design of the questions followed design principles of multiple choice questions that include the competitiveness of the distractors, aligning them with common misunderstandings and constructing other distractors similar to the correct answer and those corresponding to the misunderstandings~\cite{doi:10.3102/0034654317726529}. The nature of the questions allows for the design of distractors using logical rules. For example, a Hadamard gate can be applied to the wrong qubit to create a distractor. Then, two more distractors with similar characteristics to the two existing answer options are created such that no features stand out. The questions were reviewed internally by experts. An example of the second type of question is shown in Figure \ref{fig:pr_comp}. All \ac{rc} test items can be found in the supplementary material.

\begin{figure*}[ht]
\begin{tcolorbox}[colback=white, colframe=black]
The Hadamard gate splits the state $\ket{0}$ into equal parts, and in reverse, it transforms equal amplitudes to the state $\ket{0}$. When acting on the state $\ket{1}$, it does the same with a $\pi$ (180$^\circ$) relative phase shift. Consider the following state.\\
Note: basis states are written as $|$qubit\#3 qubit\#2 qubit\#1$\rangle$.\\
\includegraphics[trim=0 30 0 37,clip,width=0.28\linewidth]{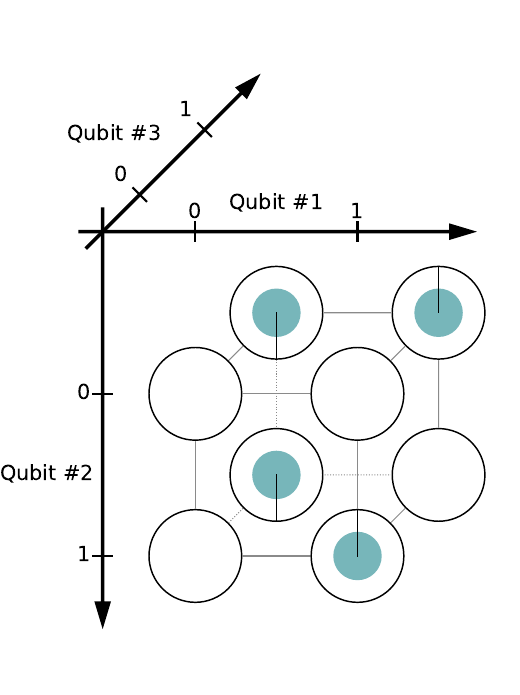}\\
What is the resulting state if a Hadamard gate is applied on qubit \#3?

\begin{minipage}[t]{0.45\textwidth}
\begin{tcolorbox}[boxsep=0pt,colback=white, colframe=d-blue, boxrule=0.5pt]
\includegraphics[trim=0 30 0 37,clip,width=0.7\textwidth]{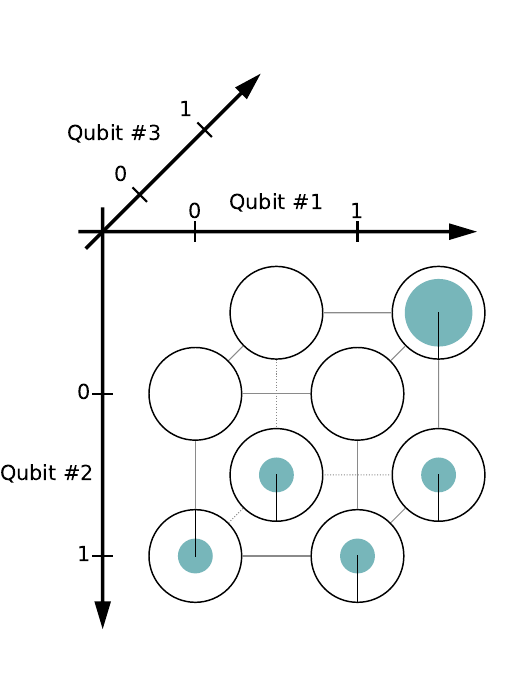}
\end{tcolorbox}

\begin{tcolorbox}[boxsep=0pt,colback=white, colframe=d-blue, boxrule=0.5pt]
\includegraphics[trim=0 30 0 37,clip,width=0.7\textwidth]{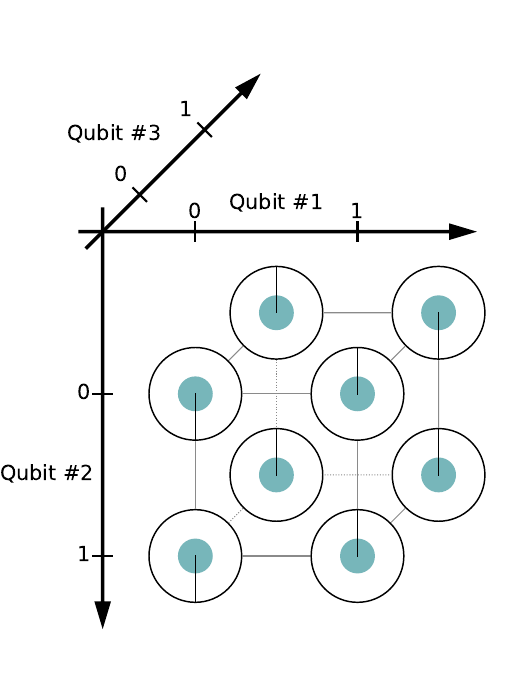}
\end{tcolorbox}
\end{minipage}
\hfill
\begin{minipage}[t]{0.45\textwidth}
\begin{tcolorbox}[boxsep=0pt,colback=white, colframe=d-blue, boxrule=0.5pt]
\includegraphics[trim=0 30 0 37,clip,width=0.7\textwidth]{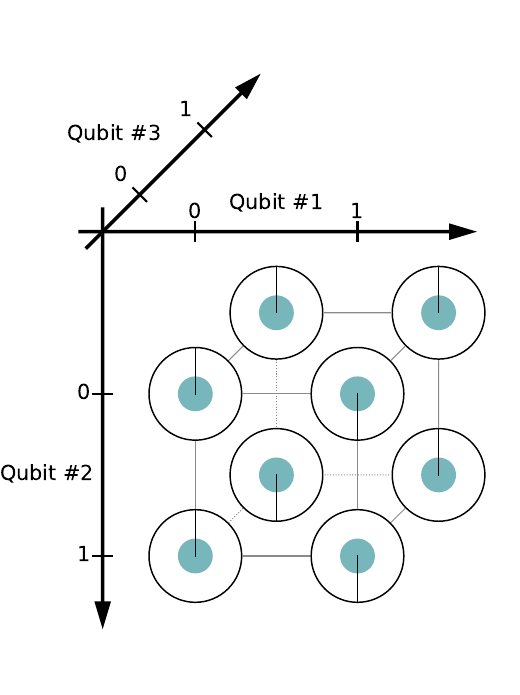}
\end{tcolorbox}

\begin{tcolorbox}[boxsep=0pt,colback=white, colframe=d-blue, boxrule=0.5pt]
\includegraphics[trim=0 30 0 37,clip,width=0.7\textwidth]{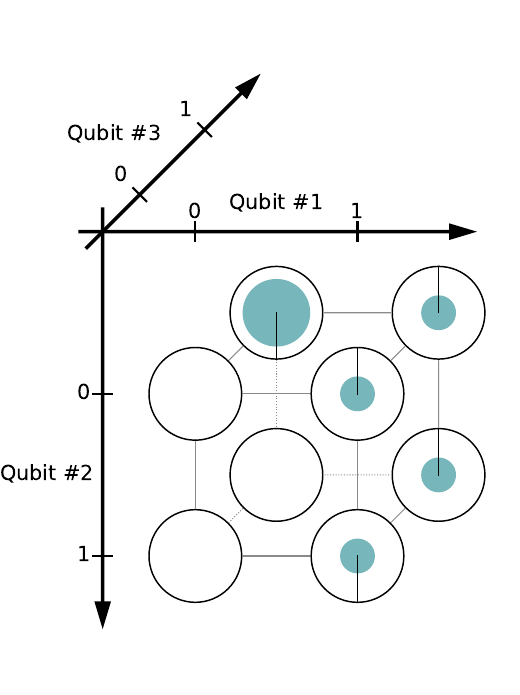}
\end{tcolorbox}
\end{minipage}
\end{tcolorbox}
\caption[Example of a procedural understanding question.]{Example of a procedural understanding question, where superpositions are destroyed. The correct answer is at the bottom left. The distractors are chosen such that the Hadamard gate is applied to qubit \#1 (top left), and with two other distractors that share features with this distractor and the correct answer.}
\label{fig:pr_comp}
\end{figure*}

\subsubsection{Hadamard gate performance assessment questions}

\begin{figure*}[ht]
\begin{tcolorbox}[colback=white, colframe=black]

\begin{minipage}[ht]{0.5\textwidth}
    Consider the following state.\linebreak
Note: basis states are written as $|$qubit\#3 qubit\#2 qubit\#1$\rangle$
\begin{equation*}
    \frac{1}{\sqrt{6}}\ket{000}-\frac{1}{\sqrt{6}}\ket{001}-\frac{1}{\sqrt{6}}\ket{010}+\frac{1}{\sqrt{6}}\ket{011}-\frac{1}{\sqrt{6}}\ket{101}+\frac{1}{\sqrt{6}}\ket{110}
\end{equation*}
What is the resulting state if a Hadamard gate is applied on qubit \#1?
\end{minipage}
\begin{minipage}[ht]{0.4\textwidth}
\includegraphics[trim=0 30 0 37,clip,width=0.62\linewidth]{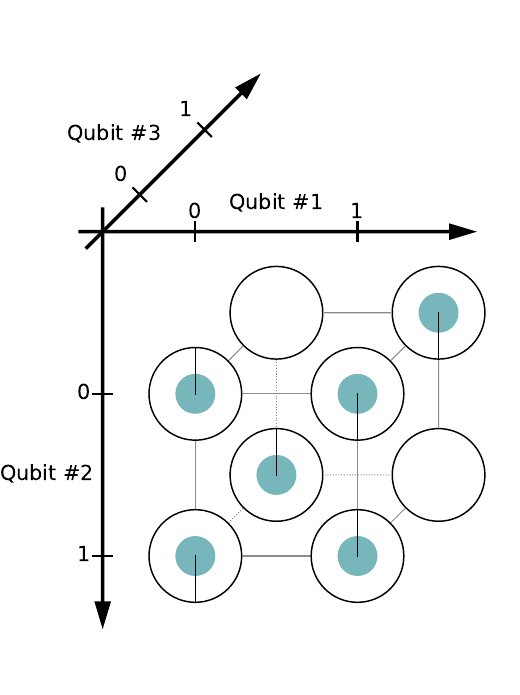}\linebreak
\end{minipage}

\begin{minipage}[t]{0.45\textwidth}
\begin{tcolorbox}[boxsep=0pt,colback=white, colframe=d-blue, boxrule=0.5pt]
\begin{equation*}
\begin{aligned}
        &\frac{1}{\sqrt{3}}\ket{001}-\frac{1}{\sqrt{3}}\ket{011}-\frac{1}{2\sqrt{3}}\ket{100}\\+ &\frac{1}{2\sqrt{3}}\ket{101}+\frac{1}{2\sqrt{3}}\ket{110}+\frac{1}{2\sqrt{3}}\ket{111}
\end{aligned}
\end{equation*}
\includegraphics[trim=0 30 0 37,clip,width=0.57\textwidth]{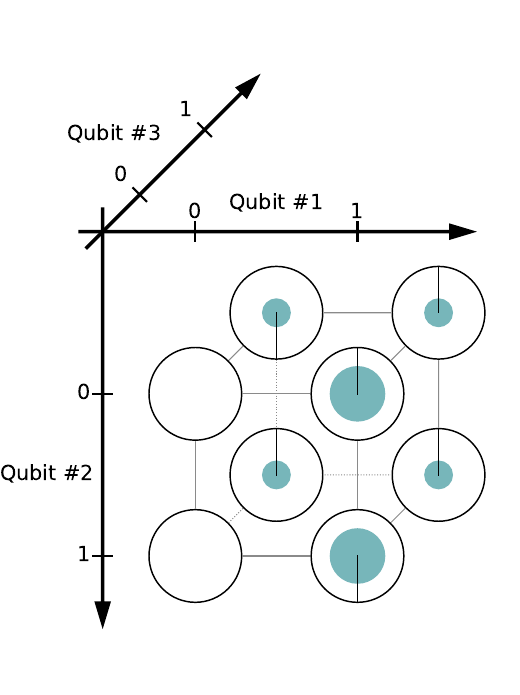}
\end{tcolorbox}
\begin{tcolorbox}[boxsep=0pt,colback=white, colframe=d-blue, boxrule=0.5pt]
\begin{equation*}
\begin{aligned}
    &\frac{1}{2\sqrt{3}}\ket{000}-\frac{1}{\sqrt{3}}\ket{001}+\frac{1}{2\sqrt{3}}\ket{011} \\ - &\frac{1}{2\sqrt{3}}\ket{100}-\frac{1}{\sqrt{3}}\ket{110}+\frac{1}{2\sqrt{3}}\ket{111}
\end{aligned}
\end{equation*}
\includegraphics[trim=0 30 0 37,clip,width=0.57\textwidth]{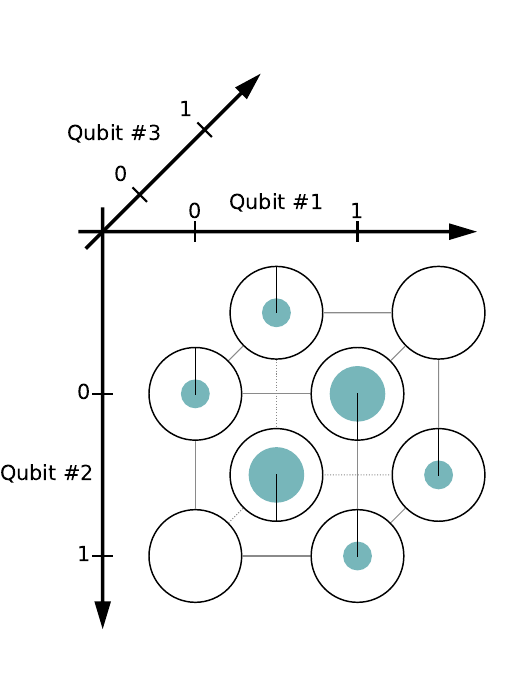}
\end{tcolorbox}
\end{minipage}
\hfill
\begin{minipage}[t]{0.45\textwidth}
\begin{tcolorbox}[boxsep=0pt,colback=white, colframe=d-blue, boxrule=0.5pt]
\begin{equation*}
\begin{aligned}
        &\frac{1}{2\sqrt{3}}\ket{001}+\frac{1}{\sqrt{3}}\ket{011}+\frac{1}{2\sqrt{3}}\ket{100}\\ -&\frac{1}{2\sqrt{3}}\ket{101}-\frac{1}{\sqrt{3}}\ket{110}+\frac{1}{2\sqrt{3}}\ket{111}
\end{aligned}
\end{equation*}
\includegraphics[trim=0 30 0 37,clip,width=0.57\textwidth]{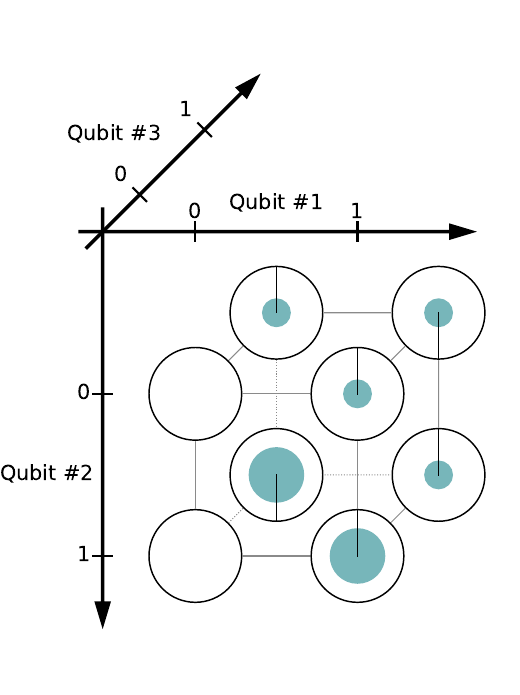}
\end{tcolorbox}
\begin{tcolorbox}[boxsep=0pt,colback=white, colframe=d-blue, boxrule=0.5pt]
\begin{equation*}
\begin{aligned}
        &\frac{1}{\sqrt{3}}\ket{000}+\frac{1}{\sqrt{3}}\ket{001}-\frac{1}{2\sqrt{3}}\ket{100}\\ -&\frac{1}{2\sqrt{3}}\ket{101}+\frac{1}{2\sqrt{3}}\ket{110}+\frac{1}{2\sqrt{3}}\ket{111}
\end{aligned}
\end{equation*}
\includegraphics[trim=0 30 0 37,clip,width=0.57\textwidth]{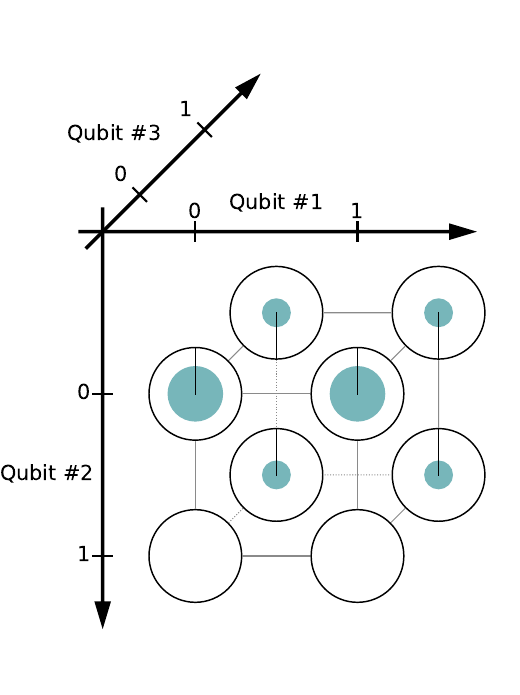}
\end{tcolorbox}
\end{minipage}
\end{tcolorbox}
\caption[Example of a main phase question with visualization.]{Example of a main phase question with visualization, where superpositions are created and destroyed. The correct answer is at the top left. The distractors are chosen such that the Hadamard gate is applied to qubit \#3 (bottom left), and with two other distractors that share features with this distractor and the correct answer.}
\label{fig:main_example}
\end{figure*}

During the main phase of the study, participants were asked questions on the outcome of a Hadamard operation on quantum states of two- or three-qubit systems, with and without visualization. As in the second type of \ac{rc} questions, in each part, participants were asked questions about the creation of superpositions, the destruction of superpositions, and a mixture of both, making it a total of six questions for each part. The creation of questions followed the same principles as the procedural understanding questions. After each question, students were asked about their level of confidence in their answer, from 0 - random guess, to 5 - very sure. During these questions, the eye movements of the participants were tracked. An example question is shown in Figure \ref{fig:main_example}. Performance was measured as the average correctness. In all cases, the answers that the participants self-reported as guesses were treated as wrong answers. All items of the \ac{rc} test can be found in the supplementary material.

\subsubsection{Cognitive load}

After part A and part B of the main phase of the study, cognitive load was measured using the instrument proposed in~\cite{Klepsch2017}, but rewritten to apply to the context of testing. Possible answers were on a Likert scale from 0 - completely wrong to 6 - absolutely right. The items are shown in the Appendix \ref{sec:A_CL}.

\subsubsection{Spatial Reasoning: Mental Rotation Ability}

The transformations with the Hadamard gate in \ac{dcn} require some amount of \ac{sra}: They involve the splitting and merging of areas along one of two or three qubit axes and the turning of a gauge in some cases. According to the taxonomy used in~\cite{Taylor2023}, it is a dynamic 2D or 3D process that uses the intrinsic reference frame of the two or three qubit-axes. Within Taylor et al.'s framework, these are the same characteristics as inherent to 2D or 3D \acf{mra}. Therefore, one can hypothesize that participants with higher \ac{mra} benefit when presented with the visualization. As a measure of \ac{mra}, the test createdby Fehringer et al.~\cite{Fehringer+2020+138+212} was used. Here, participants had to rotate parts of a cube in a Rubik's cube-like manner, instead of the whole system as alternatives such as~\cite{Henn2018}. A test item consisted of two Rubik's cube depictions, and on each item, participants had to answer whether one Rubik's cube can be transformed into another using standard Rubik's cube operations. In that aspect, it is similar to the process used in \ac{dcn}, where parts of the visualization are moving differently than others, although possible similarities are mere hypotheses at this point. 

\subsection{Eye-Tracking}

In this study, eye-tracking is used to identify the cognitive processes associated with increased performance and changes in cognitive load with visualization. Commonly used metrics are the average fixation duration and the number of transitions between different areas of interest~\cite{ALEMDAG2018413,LIU2025105263}. We argue that the former is necessary for visualization to have any effect if it is not being used. This makes it a suitable variable to moderate performance increases. The latter is used to represent integrative cognitive processes or as a proxy for mental schemata construction~\cite{https://doi.org/10.1111/jcal.12396}. More explicitly, the duration of fixation on visualization resembles engagement with the visualization during all cognitive processes, whereas the number of transitions is a measure of integration processes~\cite{https://doi.org/10.1111/jcal.13051}. By assessing the dependency of the number of transitions on performance difference with and without visualization, we can determine whether cognitive integration strategies are associated with increases in performance. 

The tasks were displayed on a 22-inch computer screen with Full HD resolution. Eye movements were recorded with a Tobii Pro Nano Eye tracker. Fixations were identified using the Identification by Velocity Threshold (I-VT) algorithm (see \cite{10.1145/355017.355028}) with thresholds of $8500^\circ/s^2$ (acceleration) and $30^\circ/s$ (velocity). Any eye movements above these thresholds would be considered saccades and were not analyzed in this work. Fixation on the \ac{dn} following a fixation on the corresponding visualization, or vice versa, was counted as a transition. Before the learning unit, we performed a nine-point calibration for each participant and repeated it until it was sufficiently accurate.

\begin{figure*}
    \centering
    \includegraphics[width=1\linewidth]{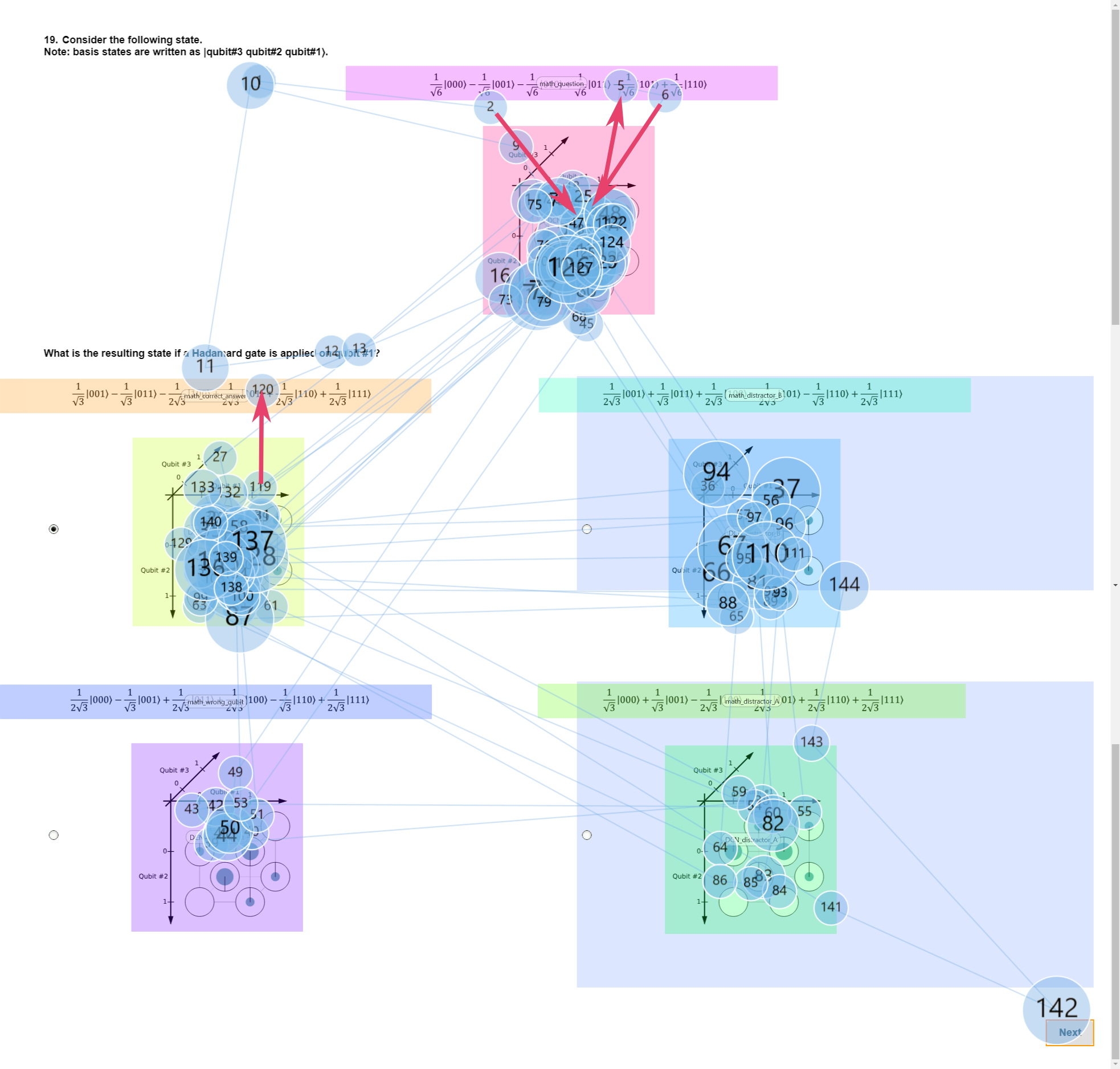}
    \caption{Eye-Tracking Areas of Interest and example gaze map of a participant solving the question shown in Figure \ref{fig:main_example}. The blue circles are numbered in order of fixation and sized according to the duration. For example, circle 87 on the correct answer in \ac{dcn} (top left) resembles a duration of $\SI{550}{\milli\second}$. The four transitions between \ac{dn} and corresponding visualization that the participant made are shown in red. The participant answered correctly, considering all answer possibilities, used almost exclusively \ac{dcn} to solve the question, and transitioned a number of times close to the minimum of the participants. The participant scored 0.5 on questions without visualization and 0.83 on questions with visualization, spending on questions with visualization on average 82\% of the time on the visualization and 18\% on \ac{dn}.}
    \label{fig:eye_tracking_aois}
\end{figure*}

The resolution of the eye-tracker is not suitable to reliably differentiate between fixations on specific features of the visualization or the symbolic notation, so \acf{aoi} were placed spanning the whole respective symbolic or visual representations. Fixations on these \ac{aoi} were counted in intervals of $\sim\SI{16.66}{\milli\second}$ and summed up to a total fixation duration for each representation, while subsequent fixations on a symbolic \ac{aoi} and the corresponding visualization (or vice-versa) were counted as transitions. Figure \ref{fig:eye_tracking_aois} shows an example of \ac{aoi} placement on the different elements of a task with a gaze map of a participant solving the question shown in Figure \ref{fig:main_example}, and the transitions that the participant made.  

\subsection{Analysis}\label{subsec:analysis}

To get a sense of overall effects and effect size, paired sample t-tests of the main metrics accuracy (one-sided), \ac{icl} (one-sided) and \ac{ecl} (two-sided) are performed and adjusted using the Holm method~\cite{doi:10.2105/AJPH.86.5.726,29def780-e117-38f0-8afb-edf384af3fad}. The Holm method is recommended in such settings, where a type I error is not a catastrophe, in favor of the more often used Bonferroni correction~\cite{STREINER201116,doi:10.2105/AJPH.86.5.726,https://doi.org/10.1111/opo.12131}. The effect sizes are determined by Cohen's d, with $0.2\leq d < 0.5$, $0.5\leq d < 0.8$, $d > 0.8$ indicating a small, medium, and large effect respectively~\cite{Sullivan2012-qq}. Then, the possible influence of \ac{mra} and the relative fixation duration on the effects of the visualization is analyzed. These influencing effects are analyzed using a linear regression with the difference in one main metric (with visualization minus without visualization). As described in~\cite{Montoya2019}, this approach is equivalent to a multiple linear regression using the score without visualization as an independent variable and the score with visualization as a dependent variable. We then analyze the effect of \ac{rc} and the number of transitions between visualization and the corresponding mathematical symbolism on the performance difference. This is methodologically equivalent to a moderator analysis, but with a focus on the slopes of the regression rather than the intercept for identification of a main effect. More technically, this means that a linear regression equation used in this work looks like

\begin{widetext}
\begin{equation}
    [\text{outcome w. vis.}] - [\text{outcome w/o vis.}] = c+m\cdot [\text{ind. variable}],
\end{equation}
\end{widetext}

describing the influence of independent variables how the visualization affects the outcome variable (accuracy or cognitive load), finding significant effects of $m$ significantly differs from 0, corrected for multiple tests. All independent variables are centered~\cite{10.3389/fpsyg.2015.00474}. 

For the paired sample t-tests, we test the normality of the differences with the Anderson-Darling test~\cite{Nelson01071998}. If the data does not look Gaussian at significance level $\alpha=0.05$, we perform a Wilcoxon signed rank test instead of the paired sample t-test~\cite{doi:https://doi.org/10.1002/0470011815.b2a15177}. For linear regression, we test for homoscedasticity with the Breusch-Pagan test~\cite{59931fb7-1472-3118-95f0-a6dba150b325} and the normality of the residuals visually using Q-Q plots~\cite{Fienberg01111979}. The results for intercept p values (moderator analyses) and slope (dependency analyses) are also adjusted using the Holm method. Unless otherwise stated, the prerequisites for the respective statistical procedure were verified.

\section{Results}\label{subsec:results}

During the main part of the study, the average time per question and standard deviation was $81\pm\SI{46}{\second}$ for questions without visualization and $72\pm\SI{32}{\second}$ for questions with visualization. For answers that were correct and not a guess, these averages were $62\pm\SI{41}{\second}$ and $81\pm\SI{31}{\second}$, respectively.

Mean answer confidence without visualization was $0.59\pm0.19$, and with visualization $0.65\pm0.20$. Mean \ac{gcl} without visualization was $0.66\pm0.16$ and mean \ac{gcl} with visualization was $0.70\pm0.17$. 

\subsection{Performance and Cognitive Load}

Table \ref{tab:main_outcome_measures} shows the results of the multicomparison t-tests of the main outcome measures accuracy, efficiency, \ac{icl} and \ac{ecl}.

\begin{table*}[ht]
    \caption{Descriptive data of the main outcome measures: mean accuracy, efficiency, \ac{icl}, and \ac{ecl} and standard deviation in parentheses, the $p$-values of the naive t-tests and the adjusted p-values, and Cohen's $d$ for the data sample of 31 participants each. Significant effects are highlighted.}
    \label{tab:main_outcome_measures}
\begin{ruledtabular}
    \begin{tabular*}{\hsize}{l*5{p{0.15\hsize}}}
        \textbf{Outcome measure} & w/o. vis. & with vis. & $t(30)$ & $p(p_{\text{adj.}})$ & Cohen's $d$\\
        \hline
        Accuracy   & $0.43\pm0.32$ & $0.53\pm0.33$ & 2.21 &  \textbf{.035 (.035)}  & 0.32 \\
        \ac{icl} & $0.76\pm0.18$ & $0.67\pm0.19$ & 2.87 & \textbf{.0037 (.0074)} & 0.48 \\
        \ac{ecl} & $0.60\pm0.21$ & $0.44\pm0.23$ & 5.02  & $\boldsymbol{<.001}$ & 0.75 
    \end{tabular*}
\end{ruledtabular}
\end{table*}

\begin{figure*}[ht]
    \centering
    \includegraphics[width=0.9\linewidth]{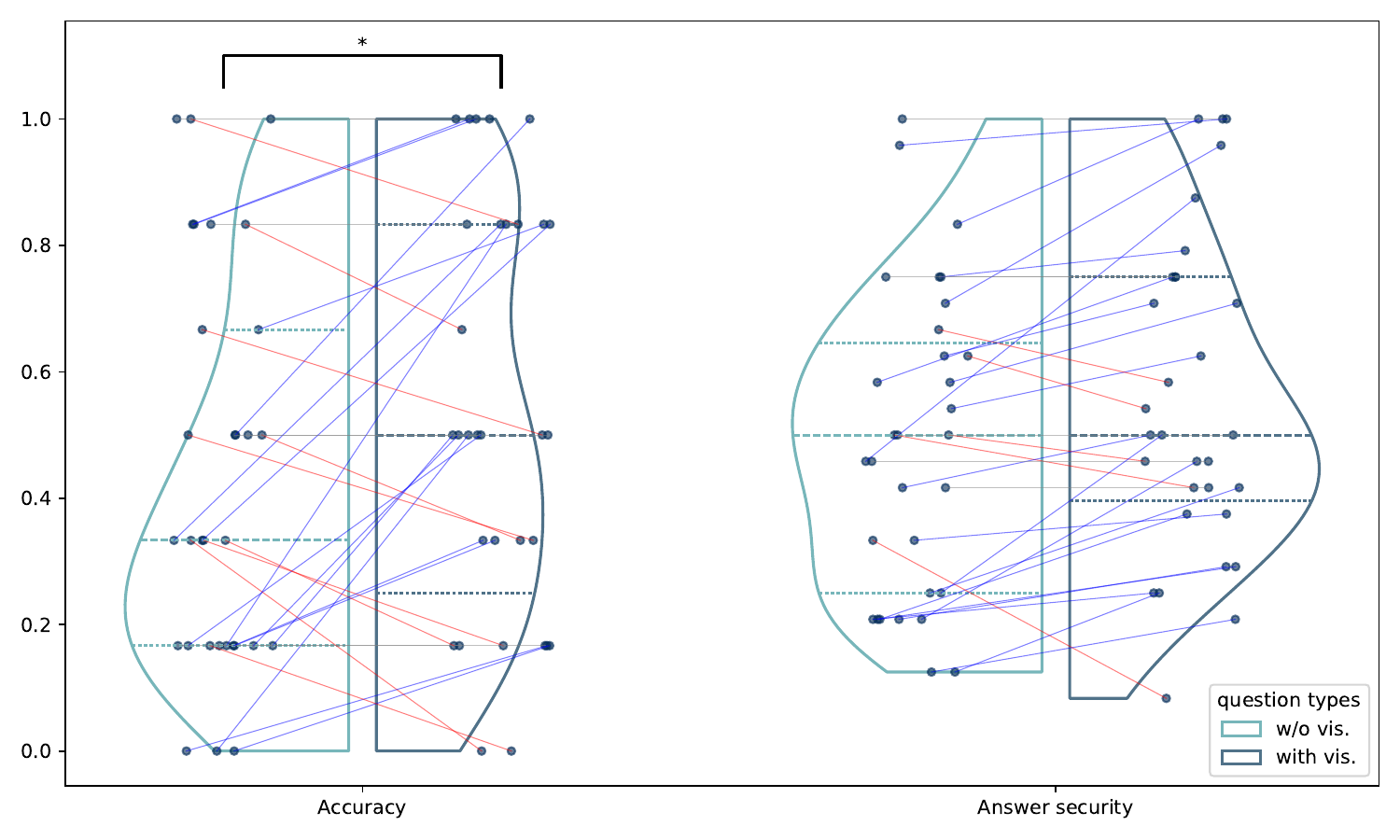}
\caption[Accuracy and answer confidence for each participant.]{Violin plots of accuracy and answer confidence (N=31). Two data points corresponding to one participant's score are connected by a line that is blue to indicate an increase, red to indicate a decrease and gray in the case of no change. The significant increase in accuracy with visualization is represented by $^{\ast}$ for $p<0.05$.}
    \label{fig:performance}
\end{figure*}

\begin{figure*}[ht]
    \centering
    \includegraphics[width=0.9\linewidth]{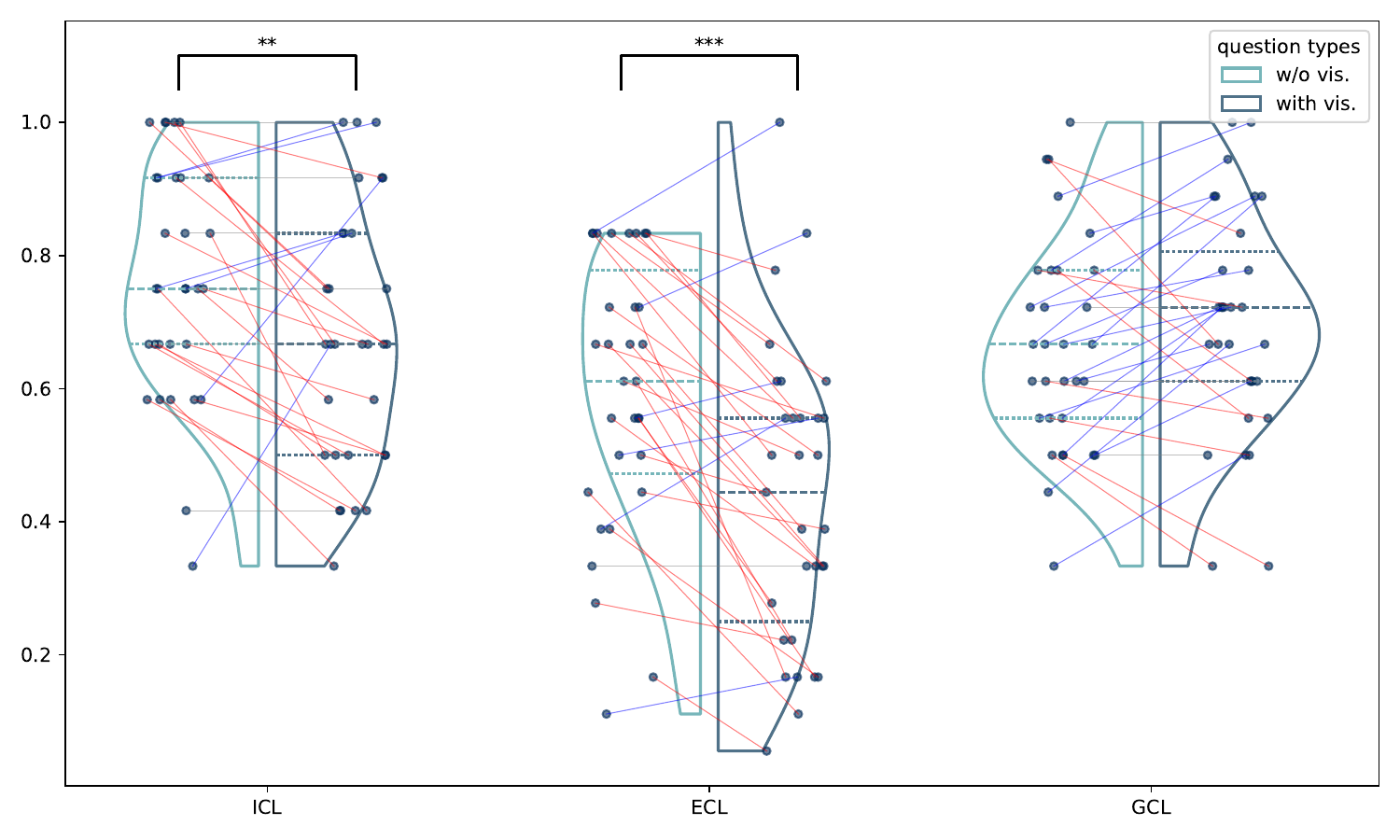}
\caption[ICL, ECL and GCL for each participant]{Violin plots of \ac{icl}, \ac{ecl} and \ac{gcl} (N=31). Two data points corresponding to one participant's score are connected by a line that is blue to indicate an increase, red to indicate a decrease and gray in the case of no change. The significant decreases in \ac{icl} and \ac{ecl} are represented by $^{\ast\ast}$ for $p<.01$ and $^{\ast\ast\ast}$ for $p<.001$, respectively.}
    \label{fig:cognitive_load}
\end{figure*}

Figure \ref{fig:performance} shows the main effects of average accuracy and answer confidence for each participant with visualization compared to without visualization. A considerable amount of participants was below the 25\% correctness threshold for guessing (10 without and with visualization). Four participants obtained a perfect score without visualization and eight with visualization. Figure \ref{fig:cognitive_load} shows the same for \ac{icl}, \ac{ecl}, and \ac{gcl}.

\subsection{Moderator analysis}

\subsubsection{Fixation duration on visualization}

We considered the relative time spent fixating on the visualization relative to the total time spent on visualization and mathematical notation for eye-tracking results of N=30 participants. On questions with visualization, the average time spent fixating on the symbolic notation was $9.2\pm\SI{10.1}{\second}$ and the average time spent fixating on the visualization was $43\pm\SI{19}{\second}$, so relatively, time spent fixating on the visualization was $0.82\pm0.18$. The average relative time spent fixating on the correct answer was $0.18\pm0.08$ when only \ac{dn} was shown and $0.15\pm0.06$ with visualization.

The coefficients of the statistical analysis of the dependency of outcome measures on the relative fixation duration on visualization are shown in Table \ref{tab:fixation_dur}. No significant effects were found.

\begin{table*}[ht]
    \caption{Analysis of the moderation effect of the relative fixation duration on the visualization based on a sample of 30 participants included in the analysis. The regression $\bigl([\text{outcome w. vis.}] - [\text{outcome w/o vis.}] = c+m\cdot [\text{fix. dur. on vis.}]\bigr)$ was tested for the outcome measures accuracy, \ac{icl}, and \ac{ecl}, with a focus on the intercept $c$. The relevant significant effects are highlighted.}
    \label{tab:fixation_dur}
\begin{ruledtabular}
    \begin{tabular*}{\hsize}{l*6{p{0.15\hsize}}}
        \textbf{Measure} & $F(1,28)$  & $R^2 (R^2_{\text{adj.}})$ & $c$ & $P(c\neq0)(P_{\text{adj.}})$ & $m$ & $P(m\neq0)$ \\
        \hline
        Accuracy & 3.32 & 0.11(0.07) & $0.11\pm0.05$ & \textbf{.02 (.02)} & $0.47\pm0.26$ & .08 \\
        \ac{icl} & 0.29 & 0.01($-0.025$) & $-0.09\pm0.03$ & \textbf{.01 (.02)} & $-0.10\pm0.18$ & .60 \\
        \ac{ecl} & 1.60 & 0.05(0.02) & $-0.16\pm0.03$ & $\boldsymbol{<.001}$ & $-0.23\pm0.18$ & .22
    \end{tabular*}
\end{ruledtabular}
\end{table*}

\subsubsection{Mental Rotation Score}

The average \acf{mra} measured for 29 participants was $0.72\pm0.21$ and the average accuracy was $0.82\pm0.13$. The average time per question was $9.8\pm\SI{4.1}{\second}$ and the average time for correct answers that were not guesses was $9.7\pm\SI{4.5}{\second}$. We analyzed the impact of the \ac{mra} on performance and cognitive load. The \ac{mra} was centered around the mean. Two participants left before taking the mental rotation test. The results of the linear regressions are shown in Table \ref{tab:mental_rot}. No significant effects were found.

\begin{table*}[ht]
    \caption{Statistical analysis of the dependence of differences in outcome measures on the \ac{mra} (N=29). The regression $\bigl([\text{outcome w. vis.}] - [\text{outcome w/o vis.}] = c+m\cdot [\text{\ac{mra}}]\bigr)$ was tested for the outcome measures accuracy, efficiency, \ac{icl}, and \ac{ecl}. The relevant significant effects are highlighted.}
    \label{tab:mental_rot}
    \begin{ruledtabular}
    \begin{tabular*}{\hsize}{l*6{p{0.15\hsize}}}
        \textbf{Measure} & $F(1,27)$  & $R^2 (R^2_{\text{adj.}})$ & $c$(SD) & $P(c\neq0)(P_{\text{adj.}})$ & $m$ & $P(m\neq0)$ \\
        \hline
        Accuracy & 1.05  & 0.037(0.002) & $0.10\pm0.05$ & \textbf{.045(.04)} & $0.39\pm0.39$ & .32 \\
        \ac{icl} & 0.58 & 0.021(-0.015) & $-0.20\pm0.26$ & \textbf{.008(.016)} & $0.20\pm0.26$ & .45 \\
        \ac{ecl} & 0.06 & 0.002(-0.035) & $-0.16\pm0.03$ & $\boldsymbol{<.001}$ & $0.06\pm0.27$ & .82 
    \end{tabular*}
\end{ruledtabular}
\end{table*}

\subsubsection{Representational Competence}

\begin{figure}[ht]
    \centering
    \includegraphics[width=\linewidth]{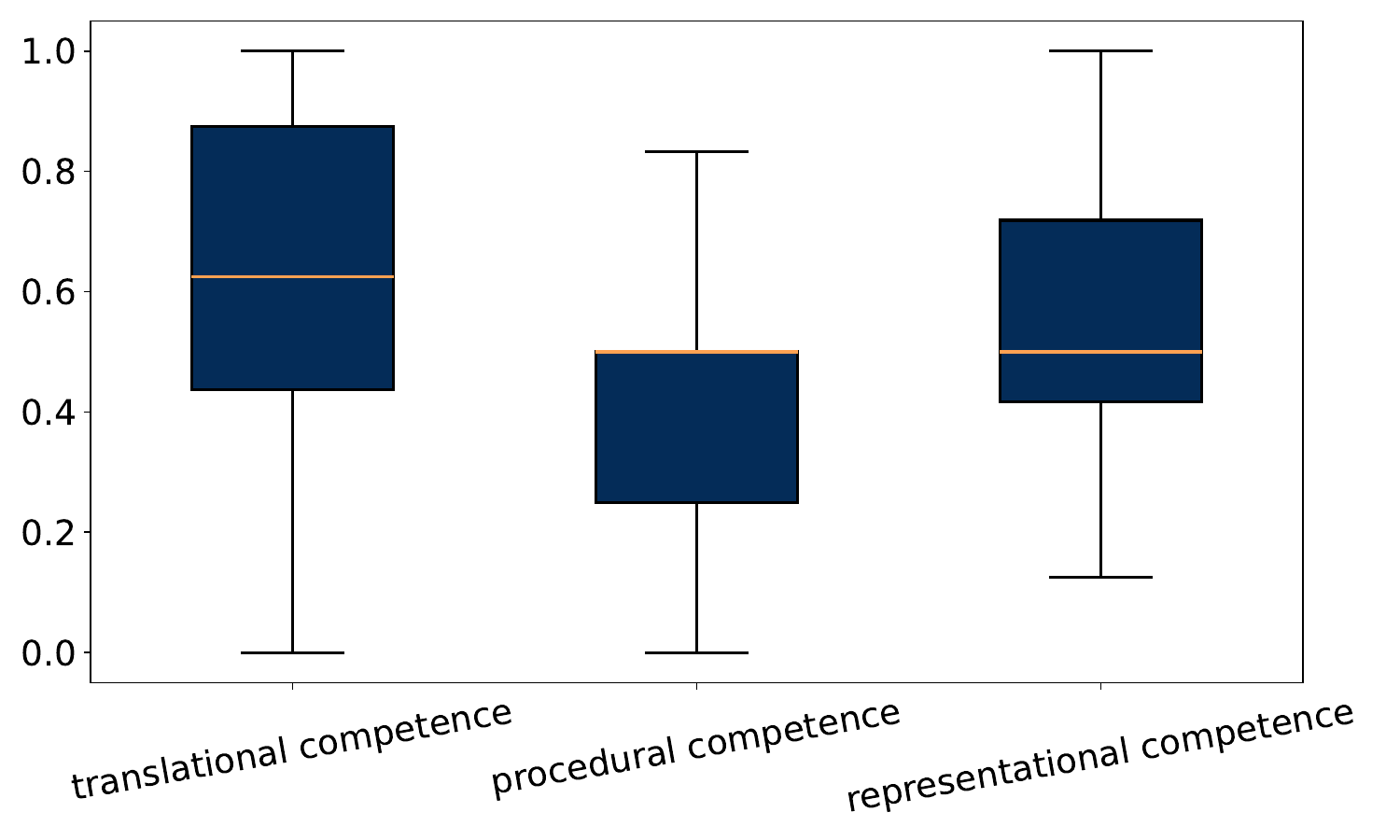}
\caption[Boxplots of the test results of different types of \ac{rc}.]{Boxplots of the test results of different (sub-)types of representational competencies (N=31). Representational competence is the average of translational competence and procedural competence.}
    \label{fig:rep_comp}
\end{figure}

Descriptive statistics on \acf{rc} are displayed in Figure \ref{fig:rep_comp}. The average procedural understanding was $0.44\pm0.26$, the average translational understanding $0.63\pm0.30$, and they combined to an average visual understanding of $0.54\pm0.23$. 

To analyze the effect of \ac{rc}, we consider the dependency of changes in outcome measures with visualization on the measured \ac{rc}. Table \ref{tab:rep_comp_stats} shows the corresponding test statistics and regression coefficients. 

\begin{table*}[ht]
    \caption{Statistical analysis of the dependence of changes in performance and cognitive load on \ac{rc} (N=31). The regression $\bigl([\text{outcome w. vis.}] - [\text{outcome w/o vis.}] = c+m\cdot [\text{\ac{rc}}]\bigr)$ was tested for the outcome measures accuracy, \ac{icl}, and \ac{ecl}. The relevant significant effects are highlighted.}
    \label{tab:rep_comp_stats}
    \begin{ruledtabular}
    \begin{tabular*}{\hsize}{l*6{p{0.15\hsize}}}
        \textbf{Measure} & $F(1,29)$  & $R^2 (R^2_{\text{adj.}})$ & $c$ & $P(c\neq0)$ & $m$ & $P(m\neq0)(P_{\text{adj.}})$ \\
        \hline
        Accuracy & 0.005  & 0.000(-0.034) & $0.10\pm0.05$ & .04 & $-0.02\pm0.21$ & .94(1.00) \\
        \ac{icl} & 0.18 & 0.006(-0.028) & $-0.09\pm0.03$ & .008 & $0.06\pm0.14$ & .68(1.00) \\
        \ac{ecl} & 7.39 & 0.20(0.18) & $-0.16\pm0.03$ & $<0.001$ & $-0.36\pm0.13$ & \textbf{.01(.03)}
    \end{tabular*}
\end{ruledtabular}
\end{table*}

The statistics for the dependency of reduction in \ac{ecl} on the \ac{rc} evaluated at $F(1,29)=6.96$, $p=0.013$, $R^2 (R^2_{\text{adj.}})= 0.19(0.17)$. The regression equation is 

\begin{equation}
    [\text{\ac{ecl} inc.}] = -(0.16\pm0.03) - (0.36\pm0.13) \cdot[\text{\ac{rc}}].
\end{equation}

\begin{figure}[ht]
    \centering
    \includegraphics[width=\linewidth]{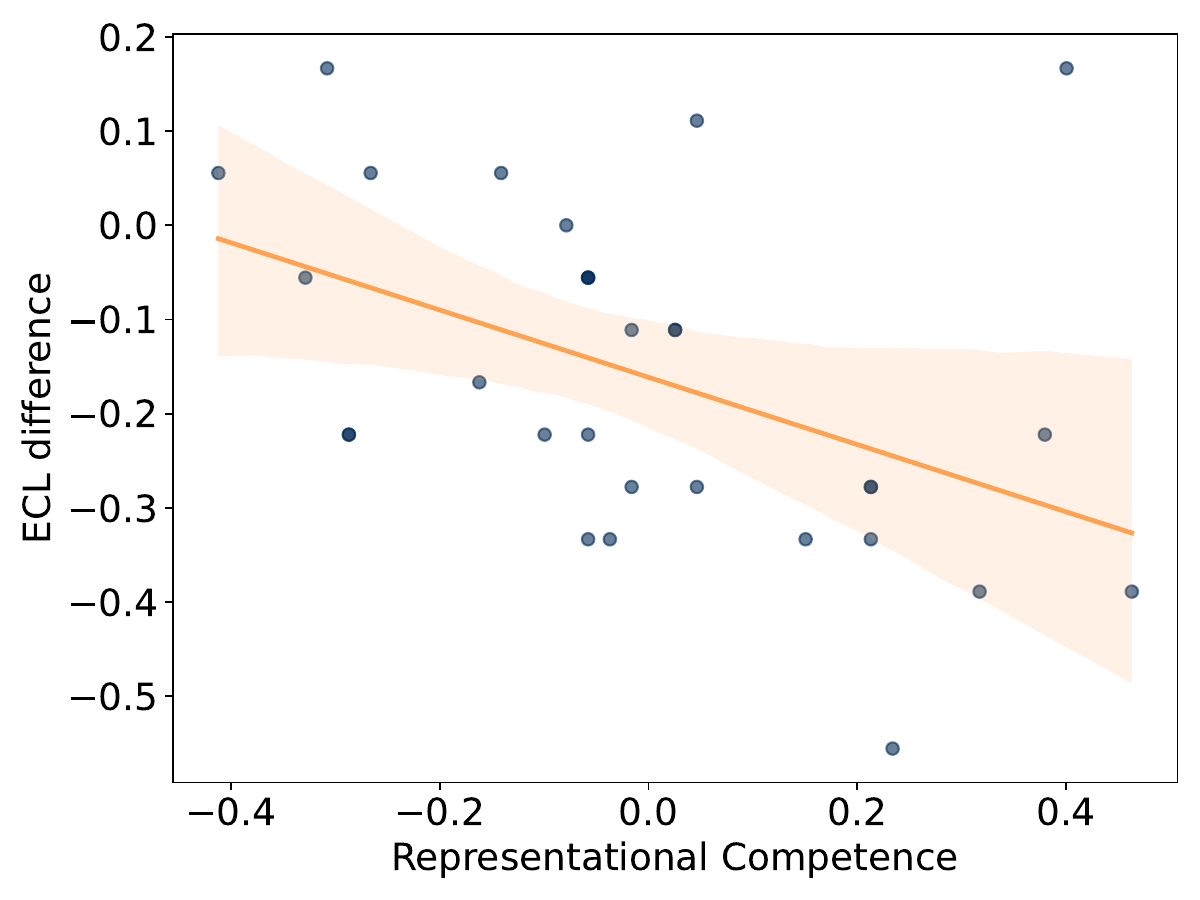}
\caption[ECL increase and the dependency on representational competence.]{Dependency of the change in \ac{ecl} with visualization on \ac{rc}, centered around the mean of $0.43\pm0.18$ (N=31). The regression equation is $\bigl([\text{ECL diff.}] = -(0.16\pm0.03) - (0.36\pm0.13) \cdot[\text{\ac{rc}}]\bigr)$. The effect is significant with $F=7.39$, $R^2 (R^2_{\text{adj.}}) = 0.20(0.18)$ and $p(p_{\text{adj.}})=0.01(0.03)$.}
    \label{fig:rc_ecl}
\end{figure}

The corresponding scatter plot is shown in Figure \ref{fig:rc_ecl}. 

\subsubsection{Number of transitions between visualization and \ac{dn}}

The maximum number of transitions between \ac{dn} and corresponding visualization per question was 40.2, and the minimum 3.3. The average total number of transitions between visualization and corresponding mathematical representation per question was $17.5\pm10.8$. The number of transitions was normalized and centered for statistical analysis. 

Significant effects were found when assessing the dependency of changes in performance with visualization on transitions between mathematics and the corresponding visualization. The coefficients of the statistical analysis of the dependence of the outcome measures on the number of transitions are shown in table \ref{tab:transitions}.

\begin{table*}[ht]
    \caption{Statistical analysis of the dependence of differences in outcome measures on the number of transitions between \ac{dn} and the corresponding visualization (N=30). The regression $\bigl([\text{outcome w. vis.}] - [\text{outcome w/o vis.}] = c+m\cdot [\text{no. of transitions}]\bigr)$ was tested for accuracy, \ac{icl}, and \ac{ecl}.}
    \label{tab:transitions}
\begin{ruledtabular}
    \begin{tabular*}{\hsize}{l*6{p{0.14\hsize}}}
        \textbf{Measure} & $F(1,28)$  & $R^2 (R^2_{\text{adj.}})$ & $c$ & $P(c\neq0)$ & $m$ & $P(m\neq0) (P_\text{adj.})$ \\
        \hline
        Accuracy & 5.41 & 0.16(0.13) & $0.11\pm0.04$ & .017 & $-0.45\pm0.19$ & \textbf{.03}(.08)  \\
        \ac{icl} & 3.61 & 0.11($0.08$) & $-0.09\pm0.03$ & .007 & 0.26(0.13) & .07(.14) \\
        \ac{ecl} & 2.43 & 0.08(0.05) & $-0.16\pm0.03$ & $<.001$ & $0.22\pm0.14$ & .13(.14)
    \end{tabular*}
    \end{ruledtabular}
\end{table*}

The statistics for the dependency of accuracy increase on the number of transitions between visualization and corresponding mathematical symbols evaluated at $F(1,28)=5.41$, $p (p_{\text{adj.}})=0.03(0.08)$, $R^2 (R^2_{\text{adj.}})= 0.16(0.13)$. The regression equation is

\begin{widetext}
\begin{equation}
    [\text{accuracy inc.}] = (0.11\pm0.04) - (0.45\pm0.19) \cdot[\text{no. of transitions}].
\end{equation}
\end{widetext}

\begin{figure}[ht]
    \centering
    \includegraphics[width=\linewidth]{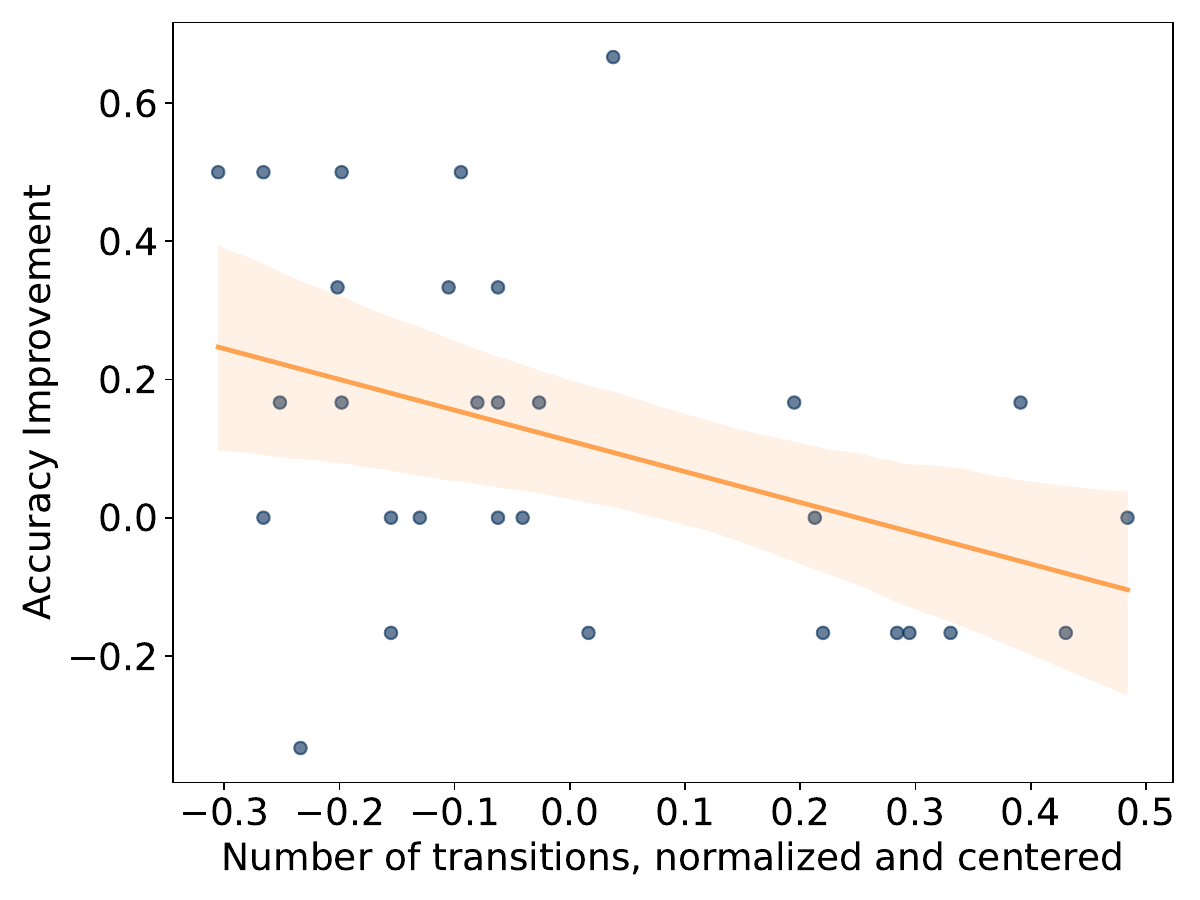}
    \caption{Dependency of accuracy improvement on the number of transitions between \ac{dn} and corresponding visualization (N=30). The number of transitions range from 3.3 to 60.2 per question, with an average of $17.5\pm10.8$ (corresponding to 0 in the figure). The regression equation is $\bigl([\text{acc. inc.}] = (0.11\pm0.04) - (0.45\pm0.19) \cdot[\text{no. of transitions}]\bigr)$. When we perform multiple-testing adjustments, the effect is not significant with $F=5.41$, $R^2 (R^2_{\text{adj.}}) = 0.16(0.13)$ and $p (p_{\text{adj.}})=0.03(0.08)$.}
    \label{fig:acc_transitions_young}
\end{figure}

Figure \ref{fig:acc_transitions_young} displays the data for the dependency of accuracy improvement on the number of transitions.

\section{Discussion}

In the following, we discuss the results in regard to our research questions, generalizability, practical applications, limitations and potential avenues for future research.

\subsection{RQ1: Impact of Visualization on Performance and Cognitive Load}

We saw a significant increase in performance (small effect size) and a significant reduction in cognitive load when visualization was present (small effect size in \ac{icl} and medium effect size in \ac{ecl}). These effects remain when controlling for \ac{mra} and fixation duration on the visualization. The findings show that the participants of two or less years of quantum physics experience benefited from the visualization. This is the first proof of the multimedia principle for testing in multi-qubit systems. The results offer an indication that the multimedia principle also holds when considering complex tasks~\cite{Hu2021,Schewior2024}.

\subsection{RQ2: Representational Competence}

We took a detailed look at the influence of \acf{rc} (visual understanding and visual fluency) and \ac{mra}. We find a significant decrease in \ac{ecl} with visualization for participants with high \ac{rc}. Previous research has shown that \ac{dn} itself is a complex representation, especially for novices~\cite{Marshman_2018}. If \ac{rc} is present, incorporating an associated visualization can help alleviate \ac{ecl}. The fact that the results gained by the incorporated test items are in line with the expected effects can be seen as an indication of a suitable \ac{rc} test instrument. However, the test instrument was not able to predict the changes in performance or \ac{icl}. It may therefore be useful to make further adjustments to the instrument used. 

\subsection{RQ3: Integration of visualization with mathematics}

We find indications that transitioning more often between \ac{dn} and the corresponding visualization was associated with decreased performance. The effect is significant without a correction for multiple tests, leading to results that should be taken with a grain of salt. The findings indicate that transitioning between \ac{dn} and \ac{dcn} less often (to a minimum of about 3 transitions per question) is related to a larger benefit from visualization in task-solving accuracy. \acl{rc} could play a role in this, as a higher number of transitions between visualization and mathematics might resemble an attempt to understand the visualization rather than solve the question. This supports the results in~\cite{https://doi.org/10.1111/jcal.13051} associating the number of transitions with horizontal coherence formation, a process of matching the visualization with mathematics that is not useful for solving the task itself.

In contrast to the results of~\cite{LINDNER201791}, we do not see participants spending more time fixating on the correct answer with visualization, indicating that the visualization might not improve the decision-making process in this context. However, participants demonstrated a higher answer confidence with visualization. Rather, considering the \ac{deft} framework, the benefits of presenting the visualization in addition to mathematics, in terms of performance and specifically \ac{icl}, probably lie in providing alternative problem-solving strategies, i.e. complementing by showing different information (the information of a qubit is stored in a different dimension in space instead of in certain basis states) and enabling different processes (the action of the Hadamard gate along an axes of a qubit instead of on each basis state individually).

\subsection{Generalizability \& Practical Implications}

It is important to consider the restricted context of this study. Although this is probably a necessity for significant findings, it leaves a lot of room for the question of how generalizable the results are. We see two possible directions of questioning the generalization of the findings: Does the utilization of different visualization techniques produce similar results? Do the findings generalize to other operations or properties of multi-qubit systems, e.g., other unitary gates, measurements, or entanglement? Do the results generalize to participant groups with more experience in quantum physics, or would we see an expertise-reversal effect?

One can argue that improving problem-solving capabilities with the Hadamard gate is of enough importance to serve as a sole justification for the use of a visualization in university courses. However, it could be that the \ac{dcn} visualization shows particular strength in this context but does not support beneficial strategies for other processes in multi-qubit systems, even also when considering the action of the Hadamard gate on states that are not classical or equal superpositions. The Bloch sphere, for example, visualizes more complex unitary operations as rotations in space which is not possible with \ac{cn}. The BEADS representation~\cite{Huber_2025}, as an extension of the Bloch sphere to multi-qubit systems that also displays entanglement properties, could therefore be a useful alternative in other contexts than the Hadamard gate on equal superpositions or ``classical states''.

To bridge considerations between task solving and learning, in practice, the performance increase with visualization might already be a considerable argument for the use of visualizations in the multi-qubit context. The decrease in cognitive load also supports this line of thinking, decreasing not only perceived difficulty of learning material by reduction of \ac{icl}, but also distracting from \ac{dn} for those students that are not as familiar with the mathematical symbolism, possibly also enhancing student motivation in the process. We also see that \ac{gcl} does not, like \ac{icl} or \ac{ecl}, decrease when visualization is presented. Therefore, we can reason that tasks become easier to solve with visualization, but do not lead to reduced effort for learning processes that \ac{gcl} is commonly argued to measure~\cite{Mayer_2021}. 

When learning \ac{qis} concepts, an important learning goal will always be to understand the underlying mathematical formalism. For this reason, despite the increase in performance and the use of visualization making tasks easier, it is advisable to show the visualization in addition to \ac{dn} for the purpose of training \ac{dn}. For performance increases and decreases in cognitive load, the results show that it is beneficial to foster \acf{rc} in students when teaching with visualizations in addition to the mathematical notation. Without this competence, students experience higher \ac{ecl} and might transition more often between mathematics and visualization in an attempt to gain \ac{rc}, perhaps missing the actual learning goal. 

In practice, visualizations such as \ac{dcn} and BEADS are most feasible in computer-based teaching instructions. While drawing such visualizations on the whiteboard or on paper is possible, it might be more tedious than writing calculations in Dirac notation. However, as today's environments largely allow for computer-based teaching, using visualization to support teaching is universally feasible. Taking into account the design principles of using \ac{mer} in learning~\cite{MAYER2021229}, we recommend the use of visualization techniques such as \ac{dcn} when designing educational material. Visualization can also be embedded into teaching in an interactive way using tools such as the one developed in \url{https://github.com/QuanTUK/} and easily accessible at \url{https://dcn.physik.rptu.de/}. In addition, the use of other representations like BEADS might lead to similar benefits and can also be used to explore new perspectives on complex quantum systems, their properties, and processes therein. 

\subsection{Limitations \& Future Research}

We were able to identify benefits of the visualization for learners with little experience in quantum physics in terms of performance. However, due to the small sample size of participants with more experience, it is impossible to examine the existence of a possible expertise-reversal effect. Students with quantum physics experience also have experience with \ac{dn}, but may have never seen the visualization before, which could result in it not providing as much of a benefit. Therefore, it remains unclear whether visualization enhances or even hinders the performance of more experienced students. With a larger participant group in this domain, a statistical analysis of the effects of years of quantum physics experience and for the more experienced participant group would be feasible.

Although we found an additional decrease of \ac{ecl} for students with higher \acf{rc} when adding a visualization to the mathematical formalism, the \ac{rc} test instrument used was unable to predict performance increases. To gain insight into how students are best taught to understand visualization, it is necessary to reevaluate the test instruments for \ac{rc} in contexts such as those examined in this study. The eye-tracking results show that it might be worth exploring the number of transitions between visualization and mathematics as an inverse measure of \ac{rc}. A problem could also be that the underlying framework~\cite{rau2017conditions} is only applicable to learning with multiple visual representations, and not to task-solving with a single visual representation. Perhaps, the decision-making step~\cite{LINDNER201791} is more important than envisioned and leads to differences in results when comparing task-solving performance and learning outcomes, an argument perhaps inspired by the higher answer confidence with visualization that we observed. 

In the present study, we focus on task solving with mathematics and one additional visualization. However, the use of more than two representations in \ac{qis} education could yield benefits that are yet to be explored in the context of multi-qubit systems~\cite{Rexigel2024}. The practical implications of such findings would be even more important due to visualizations being rather underutilized in multi-qubit systems~\cite{rexigel2024investigating}. In addition, most of the \ac{rc} framework by Rau~\cite{rau2017conditions} is not considered when only using a single visualization, more specifically, connectional understanding and meta-representational competencies.

In addition, we did not measure motivational factors that could influence visualization use. These motivational factors might lead not only to an increase in learning outcomes, but also to better task solving performance~\cite{10.3389/fpsyg.2017.01346}. As is apparent, there are many open questions that need to be answered in future research. It will be important to prioritize the questions and contexts that are especially important to the \ac{qis} education community to find a concrete way forward.

\section{Conclusion \& Outlook}

The study shows that the \ac{dcn} visualization is associated with a small increase in performance, a small reduction in \ac{icl} and a medium reduction in \ac{ecl} for participants with up to two years of experience in quantum physics, while not coinciding with a reduction in \ac{gcl}. We found indications that switching often between fixating on \ac{dn} and visualization is associated with a decrease in performance. We did not find an influence of \ac{rc} on these effects, which is reason to believe that the test instrument for \ac{rc} needs to be re-evaluated.

We investigated the effects of visualization on solving tasks with the Hadamard gate in multi-qubit systems while also exploring two connections between task solving and learning: measurement of cognitive load and tracking of eye movements. The measured cognitive load showed that visualization can significantly reduce both the unnecessary cognitive load imposed by the use of purely \ac{dn} as well as the intrinsic difficulty of the questions and, therefore, also of eventual learning material. Tracking eye movements has shown that learners who are new to the field and often transition between \ac{dn} and visualization may experience less benefit from visualization for understanding the content, although the causation here is unclear. These transitions could also be an attempt by the learners to better understand the visualization and therefore gain \ac{rc}, hinting towards the absence of such. In practice, we conclude with a recommendation to use visualization and foster \acl{rc} to ease learning of the mathematical formalism and enhance \ac{qis} education in the context of multi-qubit systems, especially for learners new to the field. 

Future studies could venture into the realm of different unitary operations, the measurement process, and identification of (partial) entanglement, with \ac{dcn} or other visualization techniques, using similar study designs. In this way, further comparisons of the benefits of different visualizations will be possible, as well as the context- and learner-dependencies of such benefits, resulting in guidelines for educators in the field of \ac{qis} as well as findings that may advance the field of educational research.

\begin{acknowledgments}
We thank the research initiative Quantum Computing for Artificial Intelligence at the RPTU Kaiserslautern-Landau for covering the study expenses.
E.R., J.B., P.L., M. K.-E., and A.W. acknowledge support by the German Federal Ministry of Education and Research (BMBF) in the QuanTUK project under grant number 13N15995. E.R. acknowledges support by the European Union via the Interreg Oberrhein programme in the project Quantum Valley Oberrhein (UpQuantVal).
This work was supported by LMUexcellent, funded by the Federal Ministry of Education and Research (BMBF) and the Free State of Bavaria under the Excellence Strategy of the Federal Government and the Länder.
\end{acknowledgments}

\section*{List of acronyms}
\begin{acronym}[itpc]
\acro{aoi}[AOI]{Areas of Interest}
\acro{clt}[CLT]{Cognitive Load Theory}
\acro{cn}[CN]{Circle Notation}
\acro{ctml}[CTML]{Cognitive Theory of Multimedia Learning}
\acro{dcn}[DCN]{Dimensional Circle Notation}
\acro{deft}[DeFT]{Design, Functions, and Tasks}
\acro{dn}[DN]{Dirac Notation}
\acro{ecl}[ECL]{Extraneous Cognitive Load}
\acro{icl}[ICL]{Intrinsic Cognitive Load}
\acro{itpc}[ITPC]{Integrated model of Text and Picture Comprehension}
\acro{gcl}[GCL]{Germane Cognitive Load}
\acro{mer}[MERs]{Multiple External Representations}
\acro{mra}[MRA]{Mental Rotation Ability}
\acro{qis}[QIS]{Quantum Information Science}
\acro{rc}[RC]{Representational Competence}
\acro{sra}[SRA]{Spatial Reasoning Ability}
\acro{stem}[STEM]{Science, Technology, Engineering, \& Mathematics}
\end{acronym}

\appendix
\section{Appendix A: Cognitive Load}\label{sec:A_CL}

Cognitive load was measured divided into \ac{icl}, \ac{ecl}, and \ac{gcl}~\cite{SWELLER1988257} using the following test items, adapted from 
\cite{Klepsch2017}. As was done in \cite{Klepsch2017}, participants were asked to rate their perceived cognitive load on a Likert scale from 0 -- completely wrong to 6 -- absolutely right.

\begin{description}
    \item[ICL 1] For the tasks, many things needed to be kept in mind simultaneously.
    \item[ICL 2] The tasks were very complex.
    \item[GCL 1] I made an effort not only to understand several details, but to understand the overall context.
    \item[GCL 2] My point while dealing with the tasks was to understand everything correct.
    \item[GCL 3] The tasks consisted of elements supporting my comprehension of the tasks.
    \item[ECL 1] During the tasks, it was exhausting to find the important information.
    \item[ECL 2] The design of the tasks was very inconvenient for problem solving. 
    \item[ECL 3] During the tasks, it was difficult to recognize and link the crucial information.
\end{description}

\end{document}